\documentclass[conference]{IEEEtran}

\usepackage{balance}
\usepackage{amsmath,amssymb,amsfonts}
\usepackage{algorithmic}
\usepackage{graphicx}
\usepackage{textcomp}
\usepackage{xcolor}
\usepackage{multirow}
\usepackage{url}
\usepackage{subcaption}
\usepackage{booktabs}
\usepackage[pagebackref=true,breaklinks=true,colorlinks,bookmarks=false]{hyperref}

\newcommand{\shadow}{\mathcal{S}}
\newcommand{\model}{\mathcal{M}}
\newcommand{\knowledge}{\mathcal{K}}
\newcommand{\attack}{\mathcal{A}}
\newcommand{\point}{\mathbf{x}_{\it Target}}
\newcommand{\train}{\mathcal{D}_{\it Train}}
\newcommand{\shadowData}{\mathcal{D}_{\it Shadow}}
\newcommand{\targetData}{\mathcal{D}_{\it Target}}

\newcommand{\featurevec}{\mathcal{X}}
\newcommand{\outputvec}{\mathcal{Y}}

\begin{document}
\title{ML-Leaks: Model and Data Independent Membership Inference Attacks and Defenses on Machine Learning Models}

\author{
\IEEEauthorblockN{
Ahmed Salem\IEEEauthorrefmark{1},
Yang Zhang\IEEEauthorrefmark{1}$^{\S}$\thanks{$^{\S}$Corresponding author}, 
Mathias Humbert\IEEEauthorrefmark{2}, 
Pascal Berrang\IEEEauthorrefmark{1},
\\
Mario Fritz\IEEEauthorrefmark{1},
Michael Backes\IEEEauthorrefmark{1}
} 
\IEEEauthorblockA{
\IEEEauthorrefmark{1}CISPA Helmholtz Center for Information Security,\\
\{ahmed.salem, yang.zhang, pascal.berrang, fritz, backes\}@cispa.saarland}
\IEEEauthorblockA{\IEEEauthorrefmark{2}Swiss Data Science Center, ETH Zurich and EPFL, 
mathias.humbert@epfl.ch}
}

\IEEEoverridecommandlockouts
\makeatletter\def\@IEEEpubidpullup{6.5\baselineskip}\makeatother
\IEEEpubid{\parbox{\columnwidth}{
    Network and Distributed Systems Security (NDSS) Symposium 2019\\    24-27 February 2019, San Diego, CA, USA\\    ISBN 1-891562-55-X\\    https://dx.doi.org/10.14722/ndss.2019.23119\\    www.ndss-symposium.org
}
\hspace{\columnsep}\makebox[\columnwidth]{}}

\maketitle

\begin{abstract}
Machine learning (ML) has become a core component of many real-world applications and training data is a key factor that drives current progress. This huge success has led Internet companies to deploy machine learning as a service (MLaaS). Recently, the first membership inference attack has shown that extraction of information on the training set is possible in such MLaaS settings, which has severe security and privacy implications.

However, the early demonstrations of the feasibility of such attacks have many assumptions on the adversary, such as using multiple so-called shadow models, knowledge of the target model structure, and having a dataset from the same distribution as the target model's training data. We relax all these key assumptions, thereby showing that such attacks are very broadly applicable at low cost and thereby pose a more severe risk than previously thought. We present the most comprehensive study so far on this emerging and developing threat using eight diverse datasets which show the viability of the proposed attacks across domains. 

In addition, we propose the first effective defense mechanisms against such broader class of membership inference attacks that maintain a high level of utility of the ML model.
\end{abstract}

\section{Introduction}

Machine learning (ML) has become a core component
of many real-world applications, ranging from image classification to speech recognition.
The success of ML has recently 
driven leading Internet companies, such as Google and Amazon,
to deploy machine learning as a service (MLaaS).
Under such services, 
a user uploads her own dataset to a server 
and the server returns a trained ML model to the user, 
typically as a black-box API.

Despite being popular,
ML models are vulnerable to various security and privacy attacks,
such as model inversion~\cite{FJR15},
adversarial examples~\cite{GSS15},
and model extraction~\cite{TZJRR16,OASF18,WG18}.
In this paper, we concentrate on one such attack, 
namely \emph{membership inference attack}.
In this setting,
an adversary aims to determine
whether a data item (also referred to as a data point) 
was used to train an ML model or not.
Successful membership inference attacks can cause severe consequences.
For instance, if a machine learning model is trained on the data collected from people with a certain disease,
by knowing that a victim's data belong to the training data of the model,
the attacker can immediately learn this victim's health status.
Previously, membership inference has been successfully conducted 
in many other domains, such as biomedical data~\cite{BBHM16} and mobility data~\cite{PTC18}.

Shokri et al.~\cite{SSSS17} present the first membership inference attack against machine learning models.
The general idea behind this attack 
is to use multiple machine learning models (one for each prediction class),
referred to as \emph{attack models}, to make membership inference
over the \emph{target model}'s output, i.e., posterior probabilities.
Given that the target model is a black-box API,
Shokri et al. propose to construct multiple \emph{shadow models}
to mimic the target model's behavior 
and derive the data necessary, 
i.e., the posteriors and the ground truth membership,
to train attack models.

There are two main assumptions made by Shokri et al.~\cite{SSSS17}.
First, the attacker needs to establish multiple shadow models
with each one sharing the same structure as the target model.
This is achieved by using the same MLaaS that trains the target model 
to build the shadow models.
Second, the dataset used to train shadow models 
comes from the same distribution 
as the target model's training data,
this assumption holds for most of the attack's evaluation~\cite{SSSS17}.
The authors
further propose synthetic data generation
to relax this assumption.
However, this approach can only 
be applied to datasets containing binary features
for efficiency reasons.

These two assumptions are rather strong
which largely reduce the scope of membership inference attacks
against ML models.
In this paper,
we gradually relax these assumptions
in order to show that far more broadly applicable attack scenarios are possible.
Our investigation shows that
indeed, membership inference in ML can be performed in an easier way 
with fewer assumptions than previously considered.
To remedy this situation,
we further propose two effective defense mechanisms.

\begin{table*}[!t]
\centering
\scalebox{1}
{
\begin{tabular}{lccc} 
\toprule
\multirow{2}{*}{Adversary type} & \multicolumn{2}{c}{Shadow model design} & Target model's\\
\cmidrule(lr){2-3}
& No. shadow models & Target model structure &  training data distribution\\
\midrule
Shokri et al.~\cite{SSSS17} & multiple & \checkmark  & \checkmark \\
Our adversary 1 & 1 & - & \checkmark \\
Our adversary 2 & 1 & - & -\\
Our adversary 3 & - & - & -\\
\bottomrule
\end{tabular}
}
\caption{An overview of the different types of adversaries.
\checkmark means the adversary needs the information
while - indicates the information is not necessary.}
\label{tab:attackoverview}
\end{table*}

\noindent\textbf{Membership Inference Attack.}
We study three different types of adversaries
based on the design and training data of shadow models.
As \autoref{tab:attackoverview} illustrates, we hereby gradually 
relax the assumptions of the previous work 
until we arrive at model and data independent adversary. 

\noindent\emph{Adversary 1.}
For the first adversary, 
we assume she has a dataset
that comes from the same distribution 
as the target model's training data.
Here, we concentrate on relaxing the assumptions on the shadow models.

We start by using only one
instead of multiple shadow models to mimic the target model's behavior.
As shadow models are established through MLaaS,
which implements the pay-per-query business model,
using one shadow model 
notably reduces the cost 
of performing the membership inference attack.

Extensive experimental evaluation 
(we use a suite of eight different datasets ranging from image to text
under multiple types of machine learning models) 
shows that
with one shadow model and one attack model,
the adversary can achieve a very similar performance
as reported by Shokri et al.~\cite{SSSS17}.
For instance, 
when the target model is a convolutional neural network (CNN) 
trained on the CIFAR-100 dataset,\footnote{\url{https://www.cs.toronto.edu/~kriz/cifar.html}}
our simplified attack achieves a 0.95 precision and 0.95 recall
while the attack with 10 shadow models and 100 attack models (as in the previous work~\cite{SSSS17})
has a 0.95 precision and 0.94 recall.

Then, we relax the assumption that
the shadow model is constructed in the same way as the target model,
In particular,
we show that training the shadow model 
with different architectures and parameters
still yields comparable attack performance.
Moreover, we propose a new approach for shadow model training,
which frees the adversary 
from even knowing the type of ML models used by the target model. 

\noindent\emph{Adversary 2.}
For this adversary,
we assume she does not 
have data coming from the same distribution as the target model's training data.
Also,
the adversary does not know the structure of the target model.
This is a more realistic attack scenario 
compared to the previous one.

We propose a \emph{data transferring attack} for membership inference
in this setting.
Concretely,
we train our single shadow model with a different dataset.
This means the shadow model here
is not used to mimic the target model's behavior
but only to capture the membership status of data points in a machine learning training set.

The main advantage of our data transferring attack
is that the adversary 
does not need to query the target model for synthetic data generation.
In contrast, the previous approach~\cite{SSSS17}
requires 156 queries on average to generate a single data point.
This means our data transferring attack is much more efficient, less costly, 
and harder to be detected by the MLaaS provider.

Experimental results show that
the membership inference attack still achieves a strong performance,
with only a few percentage drop
compared to the first adversary.
More interestingly, we show that our data transferring attack 
even works between datasets 
belonging to totally different domains.
For example,
by training a shadow model with the 20 Newsgroups text dataset,\footnote{\url{http://scikit-learn.org/stable/datasets/twenty_newsgroups.html}}
we are able to get a 0.94 precision and 0.93 recall
for attacking a target model trained on the CIFAR-100 image dataset.

\noindent\emph{Adversary 3.}
This adversary works without any shadow model,
i.e., the attack only relies on the posteriors (outcomes)
obtained from the target model when querying it with target data points. No training procedure is required at all. 
We show that statistical measures,
such as maximum and entropy,
over the target model's posteriors
can very well differentiate member and non-member data points.
To make a concrete membership inference,
we propose a threshold-choosing method.
Experiments show that 
such a simple attack 
can still achieve effective inference over multiple datasets.

All these experimental results show that membership inference
can be performed in a much simpler and more efficient way,
which further demonstrates the severe risks of ML models.

\noindent\textbf{Defense.}
To mitigate the membership risks, we propose two defense mechanisms,
i.e., \emph{dropout} and \emph{model stacking}.

\noindent\emph{Dropout.}
One reason behind membership inference attacks' effectiveness
is the inherent overfitting nature of machine learning models.
When an ML model faces a data point 
that it was trained on,
it returns a high posterior for one class compared to others.
Therefore, to defend against membership inference,
we use a classical approach adopted in deep learning,
namely dropout, which aims at preventing overfitting.
Dropout randomly deletes in each training iteration a fixed proportion of edges in a fully connected neural network model. 

Experiments on multiple datasets show that dropout can be a very effective countermeasure against membership inference.
On the CIFAR-100 dataset,
dropout (with 0.5 dropout ratio)
decreases the performance of our first adversary
from 0.95 precision and 0.95 recall
to 0.61 and 0.60, respectively.
Moreover, it almost preserves the same utility as the initial target model:
The target model's prediction accuracy 
only drops from 0.22 to 0.21 (CIFAR-100).
As dropout serves as a regularizer,
we observe that, 
for several learning problems, e.g., the Purchase-100 dataset~\cite{SSSS17}, 
the target model's accuracy even improves after applying dropout. 
Therefore, these models improve in performance {\it and} resilience in membership inference attacks.

\noindent\emph{Model Stacking.}
Although the dropout mechanism is effective, 
it is specific to deep neural networks.
For target models using other machine learning classifiers,
we propose a second defense mechanism, namely model stacking.
Model stacking is a major class of ensemble learning.
In model stacking, multiple ML models are organized in a hierarchical way
to prevent overfitting.
In our case, we construct the target model
with three different machine learning models.
Two models are placed in the first layer
directly taking the original training data as input,
while the third model is trained with the posteriors of the first two models.

Through extensive experiments,
we show that model stacking
is able to significantly reduce the membership inference's performance.
For instance, both precision and recall of the attack (adversary 1) drop by more than 30\%
on the CIFAR-100 dataset trained with model stacking.
Meanwhile, the target model's prediction performance stays almost the same.

In summary, we make the following contributions:
\begin{itemize}
\item We broaden the class of membership inference attacks by substantially relaxing the adversarial assumptions.
\item We evaluate membership privacy threat under
three different adversarial setups on eight diverse datasets, ultimately arriving at a model and data independent adversary.
Extensive experiments demonstrate the severe membership privacy threat
for machine learning models.
\item We propose two defense mechanisms, namely dropout and model stacking,
and demonstrate their effectiveness experimentally.
\end{itemize}

\noindent\textbf{Organization.}
The rest of the paper is organized as the following.
\autoref{sec:preliminaries} introduces the definition of membership inference
against ML models
and datasets used in the paper.
\autoref{sec:attack1},~\autoref{sec:attack2} and~\autoref{sec:attack3}
present the threat models, attack methodologies, and evaluations 
of our three different types of adversaries,
respectively.
In \autoref{sec:defense}, we introduce the two defense mechanisms.
\autoref{sec:relwork}
discusses the related work
and \autoref{sec:conclu} concludes the paper.

\section{Preliminaries}
\label{sec:preliminaries}

In this section, 
we first define membership inference attack in the machine learning setting.
Then, we introduce the datasets used for our evaluation.

\subsection{Membership Inference Against Machine Learning Models}

In this paper, we concentrate on machine learning classification,
as it is the most common ML application.
An ML classifier is essentially a function $\model$ 
that maps a data point $\featurevec$ (a multidimensional feature vector)
to an output vector $\model(\featurevec)=\outputvec$.
The length of $\outputvec$ is equal to the number of classes considered.
For most of the classification models,
the output vector $\outputvec$ can be interpreted 
as a set of posterior probabilities over all classes,
and the sum of all the values in $\outputvec$ is 1.
The parameters of an ML model are learned 
on a training dataset (denoted by~$\train$) containing multiple data points
following a predefined learning object.

Membership inference attack in the ML setting emerges 
when an adversary aims to find out whether her target data point
is used to train a certain ML model.
More formally,
given a target data point $\point$, a trained machine learning model $\model$,
and external knowledge of an adversary, denoted by $\knowledge$,
a membership inference attack (\emph{attack model}) 
can be defined as the following function.
\[
\attack: \point, \model, \knowledge \rightarrow \{0, 1\}
\]
Here, 0 means $\point$ is not a member of $\model$'s training dataset $\train$ 
and 1 otherwise.
The machine learning model $\model$ that the adversary targets
is also referred to as the \emph{target model}.
As in the work of Shokri et al.~\cite{SSSS17},
we assume the adversary only has black-box access to the target model, such as an MLaaS API,
i.e., the adversary can submit a data point to $\model$
and then obtain the probabilistic output, i.e., $\model(\point)$.

The attack model $\attack$ is essentially a binary classifier.
Depending on the assumptions,
it can be constructed in different ways,
which will be presented in later sections.

\subsection{Datasets Description}

We utilize 8 different datasets in this paper to conduct our experiments.
Among them, 6 datasets\footnote{We excluded \emph{Texas hospital stays} dataset~\cite{SSSS17}, 
as there is not enough information provided for preprocessing it.} 
are the same as the ones used by Shokri et al.~\cite{SSSS17},
i.e., MNIST,\footnote{\url{http://yann.lecun.com/exdb/mnist/}} 
CIFAR-10, CIFAR-100,
Location~\cite{YZQ16},
Purchase,\footnote{\url{https://www.kaggle.com/c/acquire-valued-shoppers-challenge/data}}
and Adult.\footnote{\url{https://archive.ics.uci.edu/ml/datasets/adult}}
We follow the same procedure to preprocess all these datasets.

In particular, the Purchase dataset does not contain any prediction classes.
Following Shokri et al.~\cite{SSSS17},
we adopt a clustering algorithm, namely K-means, to manually define classes.
The numbers of classes include 2, 10, 20, 50, and 100,
therefore, we extend the Purchase dataset into 5 datasets.
For instance, Purchase-100 represents the Purchase dataset with 100 different classes.

Moreover, we make use of two other datasets, namely News and Face, in our evaluation.
We briefly describe them in the following.

\noindent\textbf{News.}
The News dataset (20 Newsgroups)
is one of the most common datasets used for text classification and clustering.
The dataset consists of 20,000 newsgroup documents categorized into 20 classes.
The number of data points belonging to each class is very similar,
i.e., the dataset has a balanced class distribution.
We preprocess the News dataset
by first removing headers, footers, and quotes from the documents.
Then, we build the TF-IDF matrix out of the raw documents.

\noindent\textbf{Face.}
The Face dataset (Labeled Faces in the Wild\footnote{\url{http://vis-www.cs.umass.edu/lfw/}})
consists of about 13,000 images of human faces crawled from the web.
It is collected from 1,680 participants
with each participant having at least two distinct images in the dataset.
In our evaluation, we only consider people with more than 40 images, which leaves us with 19 people's data, i.e., 19 classes.
The Face dataset is challenging for facial recognition,
as the images are taken from the web and not under a controlled environment, such as a lab.
It is also worth noting that 
this dataset is unbalanced.
\section{Towards Model Independent Membership Inference Attacks (Adversary 1)}
\label{sec:attack1}

In this section, we describe our first adversary considered for membership inference attack.
For this adversary,
we mainly relax the assumption on her shadow model design.
In consequence, membership inference attack 
can be performed in a much more efficient and less costly way.

We start by defining the threat model.
Then, we describe our first simplification, i.e., using one shadow model instead of multiple.
In the end, we propose our second simplification
which frees the adversary from knowing the target model's structure.

\subsection{Threat Model}
\label{sec:attacker1Knowledge}

We define our attack model $\attack$ as a supervised ML classifier
with binary classes (member or non-member).
To train $\attack$,
the adversary needs to derive the labeled training data.
i.e., the ground truth membership.
As mentioned in \autoref{sec:preliminaries},
the adversary only has black-box access to the target model,
i.e., she is not able to extract the membership status from the target model.
Therefore, the adversary trains a shadow model~\cite{SSSS17}
to mimic the behavior of the target model,
and relies on the shadow model to obtain the ground truth membership
to train $\attack$.

To train the shadow model,
we assume that the adversary 
has a dataset, denoted by $\shadowData$,
that comes from the same underlying distribution 
as the training data for the target model.
Note that most of the experiments by Shokri et al.~\cite{SSSS17} make the same assumption.

We further assume that the shadow model
uses the same ML algorithm and
has the same hyperparameters as the target model.
To achieve this in practice, 
the adversary 
can either rely on the same MLaaS provider which builds the target model
or perform model extraction to approximate the target model~\cite{TZJRR16,OASF18,WG18}.
Later in this section, 
we show this assumption can be relaxed as well.

\begin{figure*}[!t]
\centering
\begin{subfigure}{0.47\textwidth}
   \includegraphics[width=\linewidth]{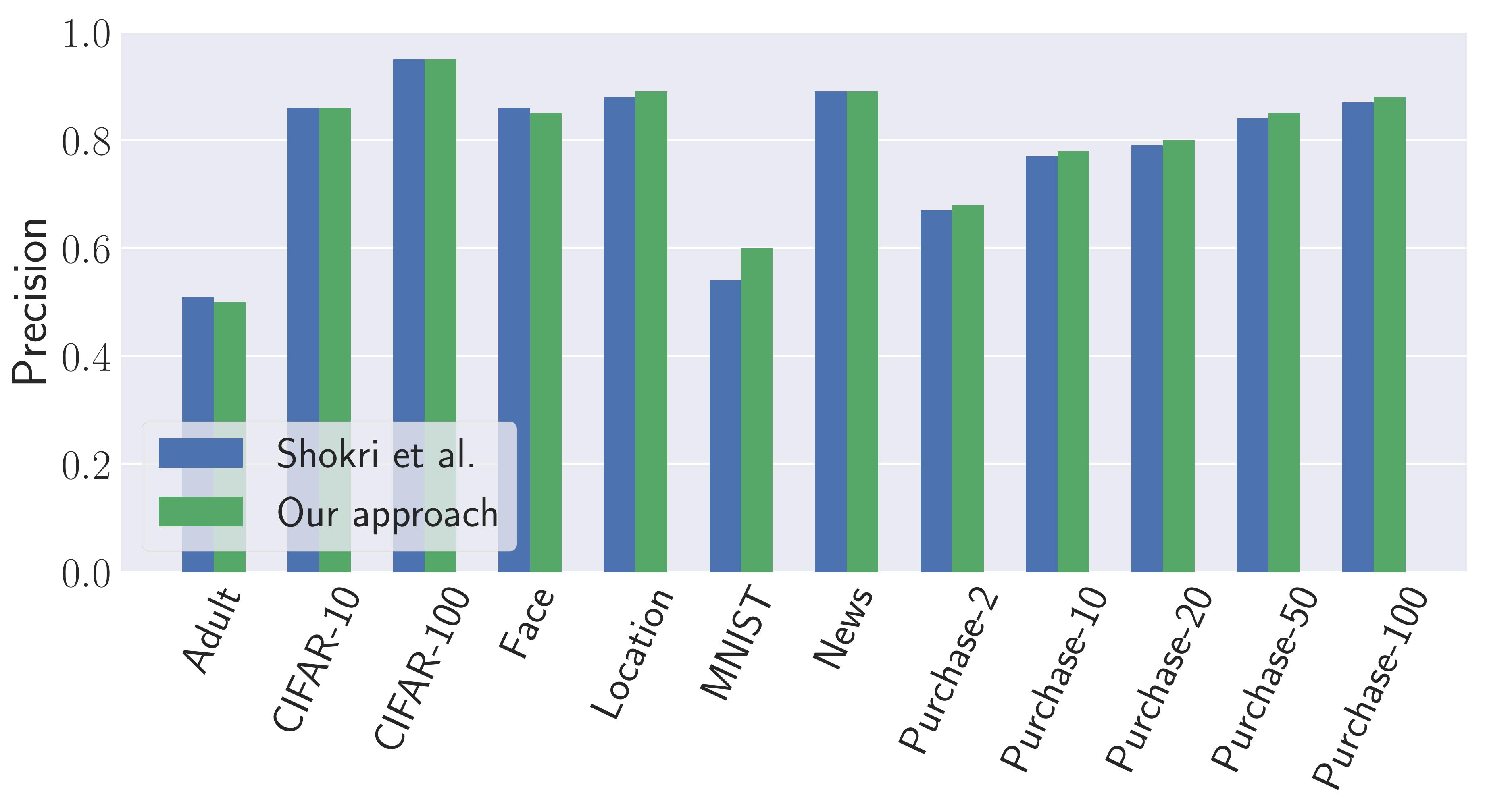}
   \caption{Precision.}
   \label{fig:shokriCompPrec} 
\end{subfigure}
\begin{subfigure}{0.47\textwidth}
   \includegraphics[width=\linewidth]{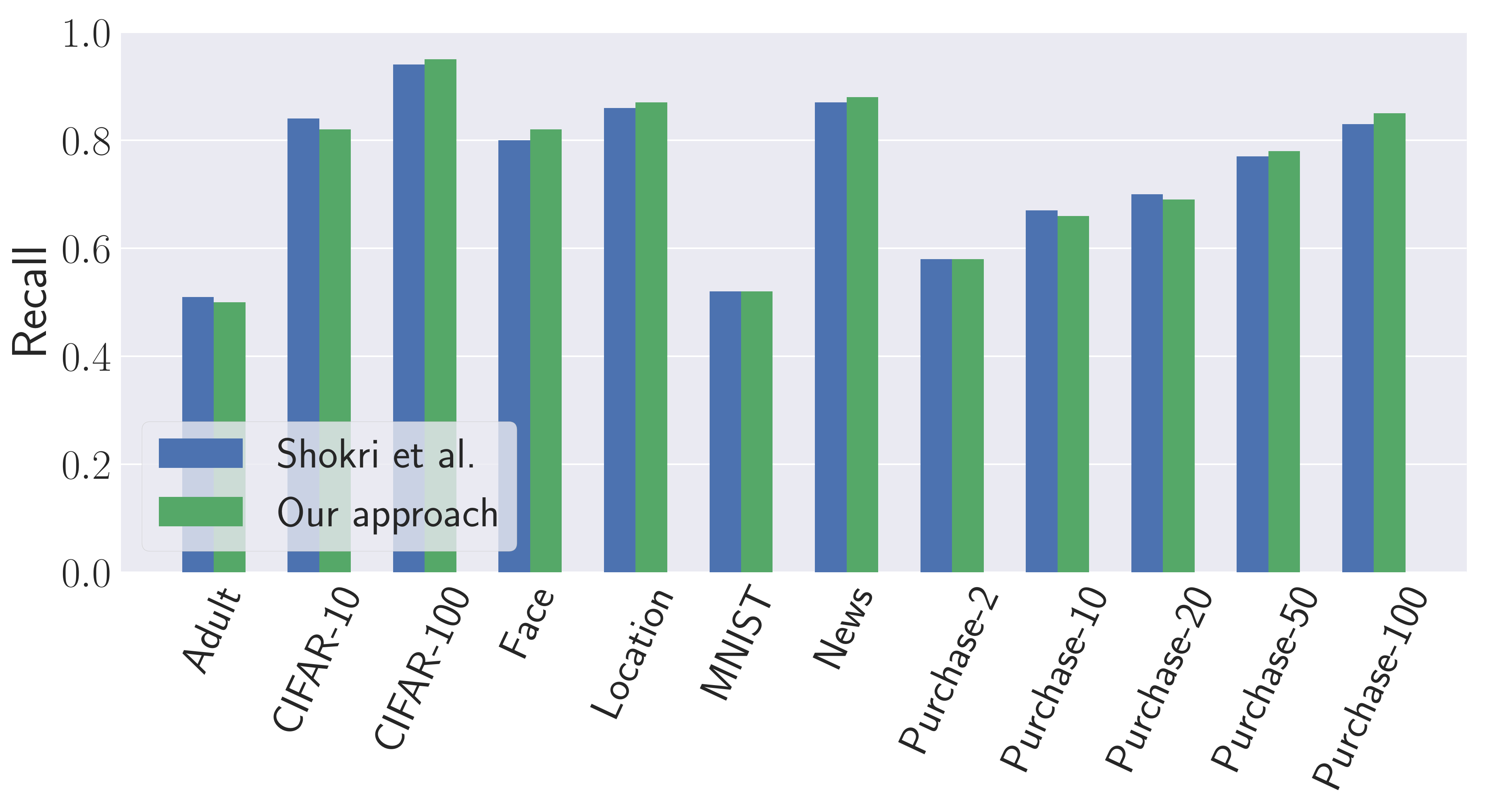}
   \caption{Recall.}
   \label{fig:shokriCompRecall}
\end{subfigure}
\caption{Comparison of the first adversary's performance with Shokri et al.'s using all datasets.
(a) precision, (b) recall.}
\label{fig:compareWithShokri}
\end{figure*}

\subsection{One Shadow Model}
\label{sec:attack1Meth}

\noindent\textbf{Methodology.}
The adversary's methodology 
can be organized into three stages,
i.e., shadow model training, attack model training, and membership inference.

\noindent\emph{Shadow Model Training.}
The adversary first splits her dataset, i.e.,~$\shadowData$, into two disjoint sets, 
namely $\shadowData^{\it Train}$ and $\shadowData^{\it Out}$.
Then, she uses $\shadowData^{\it Train}$ 
to train her only shadow model, denoted by $\shadow$.

\noindent\emph{Attack Model Training.}
The adversary uses the trained shadow model $\shadow$
to perform prediction over all data points in $\shadowData$
(consisting of $\shadowData ^{\it Train}$ and $\shadowData ^{\it Out}$),
and obtain the corresponding posterior probabilities.
For each data point in~$\shadowData$,
she takes its three largest posteriors (ordered from high to low)
or two in the case of binary-class datasets
as its \emph{feature vector}.
A feature vector is labeled as 1 (member),
if its corresponding data point is in $\shadowData^{\it Train}$,
and as 0 (non-member) otherwise.
All the generated feature vectors and labels 
are then used to train the attack model $\attack$.

\noindent\emph{Membership Inference.}
To perform the attack on whether $\point$ is in $\train$, 
the adversary queries $\model$ with $\point$ 
to obtain the corresponding posteriors.
Then, she picks the 3 maximal posteriors, again ordered from high to low,
and feed them into $\attack$ to obtain the membership prediction.

It is important to note that
our adversary only uses one shadow model 
and one attack model in her attack,
while the approach by Shokri et al.~\cite{SSSS17} 
adopts multiple shadow models
as well as multiple attack models (one for each class).
In particular, as each shadow model is established through MLaaS~\cite{SSSS17},
this strategy will largely reduce the cost of her membership inference attack.

\noindent\textbf{Experimental Setup.}
We evaluate our attack over all datasets.
For each dataset, we first split it by half into $\shadowData$ and $\targetData$.
Following the attack strategy,
we split $\shadowData$ by half into $\shadowData^{\it Train}$ and $\shadowData^{\it Out}$.
$\targetData$, on the other hand, 
is used for attack evaluation,
it is also split by half:
One is used to train the target model, i.e., $\train$,
and serves as the members of the target model's training data,
while the other serves as the non-member data points.

For image datasets, i.e., MNIST, CIFAR-10, CIFAR-100, and Face,
we use convolutional neural network (CNN) to build the target model.
Our CNN is assembled with two convolutional layers and two pooling layers
with one hidden layer containing 128 units in the end.
For the other datasets, 
we use multilayer perceptron (neural network) 
with one hidden layer (128 units) as the target model.
Each shadow model's structure is the same as its corresponding target model,
following the assumption that the adversary knows the target model's structure.
The attack model is established with another multilayer perceptron (a 64-unit hidden layer and a softmax output layer).
All our experiments are implemented in Python with Lasagne.\footnote{\url{https://github.com/Lasagne/Lasagne}}
For reproducibility purposes, our code is available at \url{https://github.com/AhmedSalem2/ML-Leaks}.

We compare our attack against the attack by Shokri et al.~\cite{SSSS17}.
Following the original configuration of the authors' code,\footnote{\url{https://github.com/csong27/membership-inference}}
we train 10 shadow models and multiple attack models (one for each class).

As membership inference is a binary classification,
we adopt precision and recall as our evaluation metrics.
Moreover, we use accuracy to measure the target model's prediction performance.

\noindent\textbf{Results.}
\autoref{fig:compareWithShokri} depicts the first adversary's performance.
In general, we observe that 
our attack has a very similar membership inference 
as the previous work~\cite{SSSS17}.
For instance,
our attack on the CIFAR-100 dataset
achieves 0.95 for both precision and recall,
while the attack by Shokri et al.\ has a 0.95 precision and 0.94 recall. 
It is also interesting to see that our attack works for both balanced datasets, such as CIFAR-10, and unbalanced datasets, such as Face.

\begin{figure}[!t]
\centering
\begin{subfigure}{0.23\textwidth}
   \includegraphics[width=\linewidth]{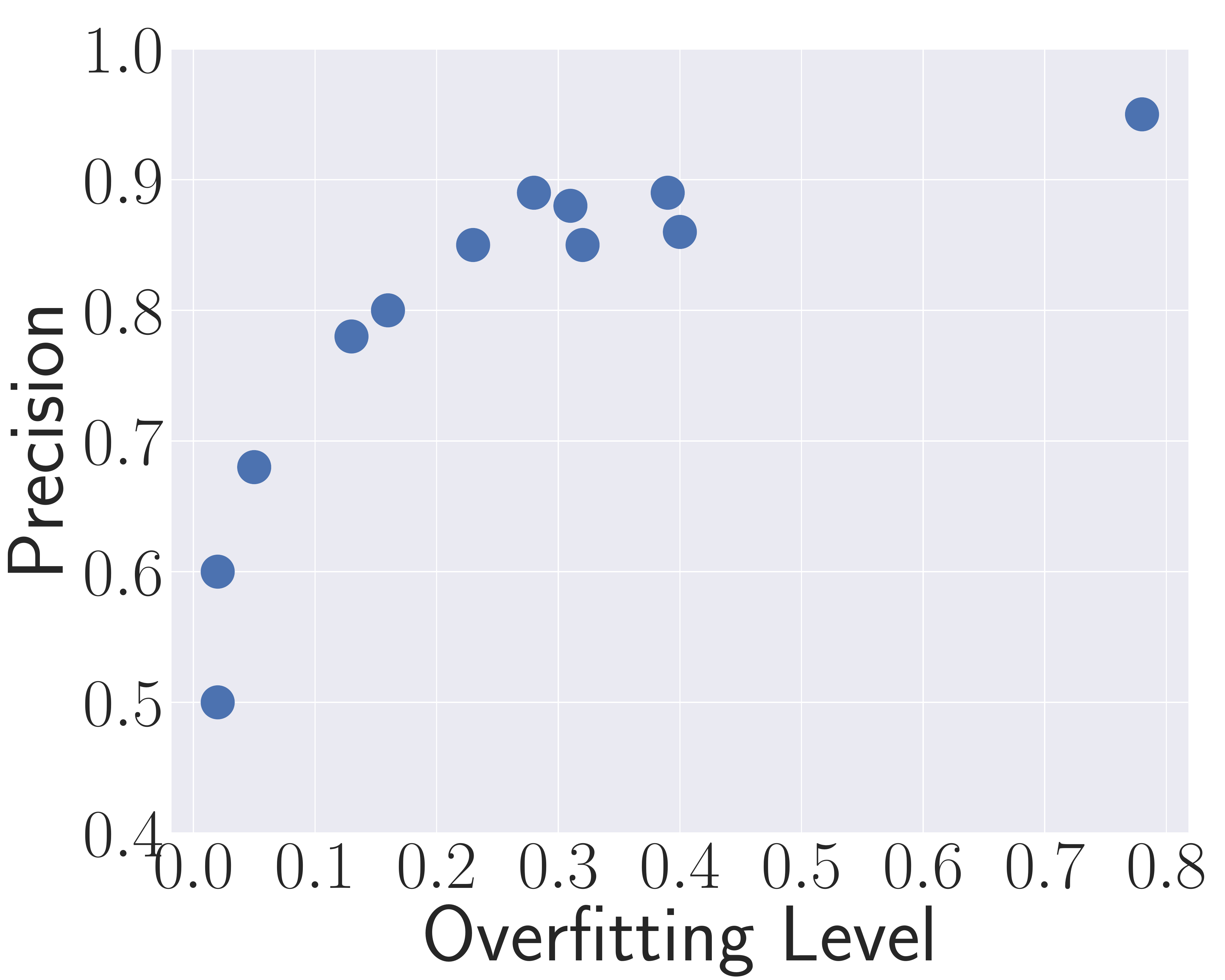}
   \caption{}
   \label{fig:gapVsPrecision} 
\end{subfigure}
\begin{subfigure}{0.23\textwidth}
   \includegraphics[width=\linewidth]{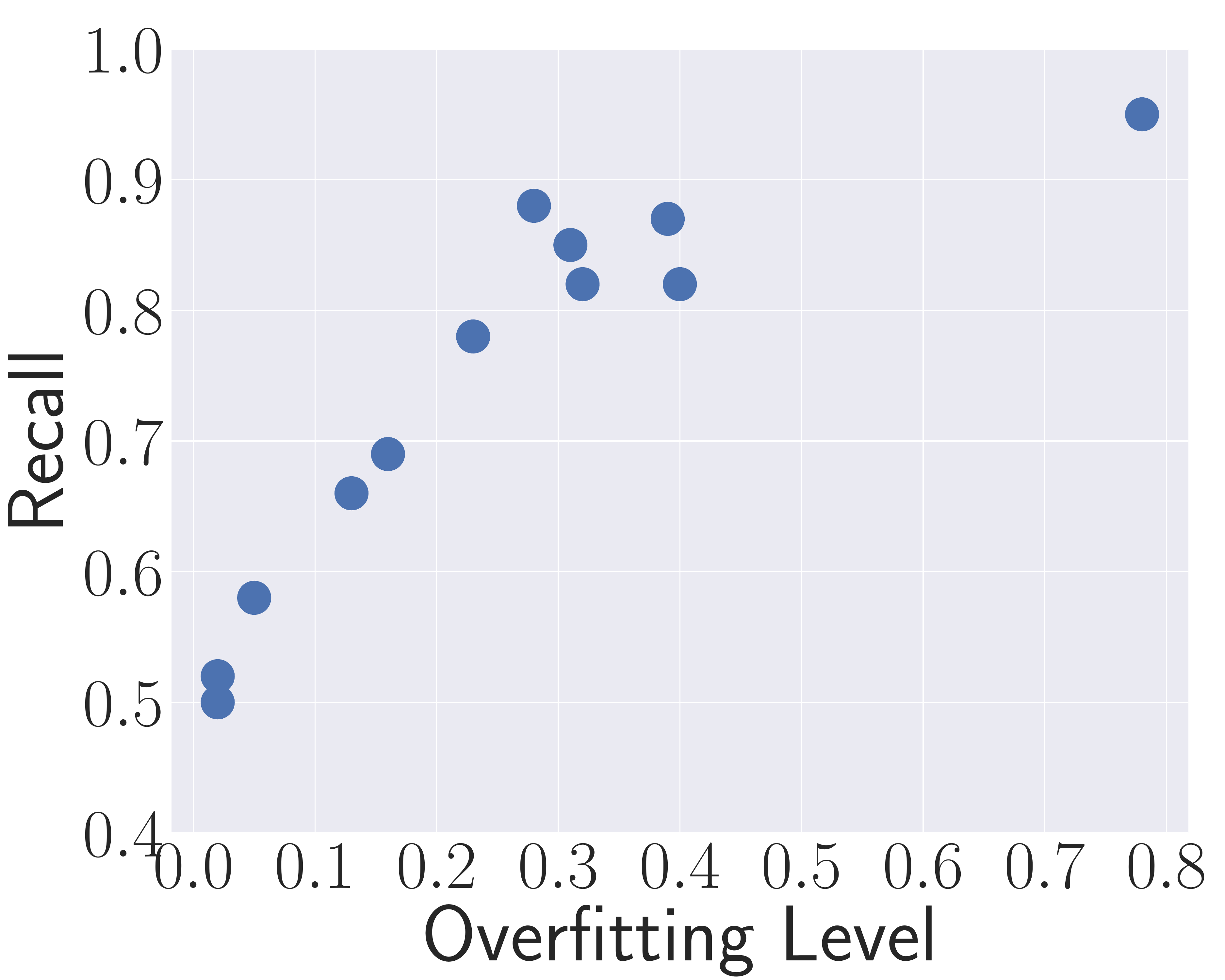}
   \caption{}
   \label{fig:gapVsRecall}
\end{subfigure}
\caption{The relation between the overfitting level of the target model
measured by the difference between prediction accuracy on training set and testing set (x-axis) and membership inference attack performance (y-axis). (a) precision, (b) recall.}
\label{fig:gapVsPerformance}
\end{figure}

We also observe variations of the attack performance on different datasets.
We relate this to the \emph{overfitting level} of ML models on different datasets, similar to previous works \cite{SSSS17,YGFJ18}.
We quantify the overfitting level of a target model as the difference between its prediction accuracy on the training set and testing set.
Through investigation,
we discover that if an ML model is more overfitted, 
then it is more vulnerable 
to membership inference attack (see~\autoref{fig:gapVsPerformance}).
For instance,
our attack on the Adult dataset achieves a relatively weak performance 
(around 0.5 precision and recall),
and there is only a 2\% difference between the target model's training and testing accuracy.
On the other hand,
the membership inference attack achieves a 0.95 precision and recall
on the CIFAR-100 dataset.
Meanwhile, the corresponding target model
provides a much better prediction performance on the training set
than on the testing set, i.e., 78\% difference.

\begin{figure}[!t]
\centering
\begin{subfigure}{0.23\textwidth}
   \includegraphics[width=\linewidth]{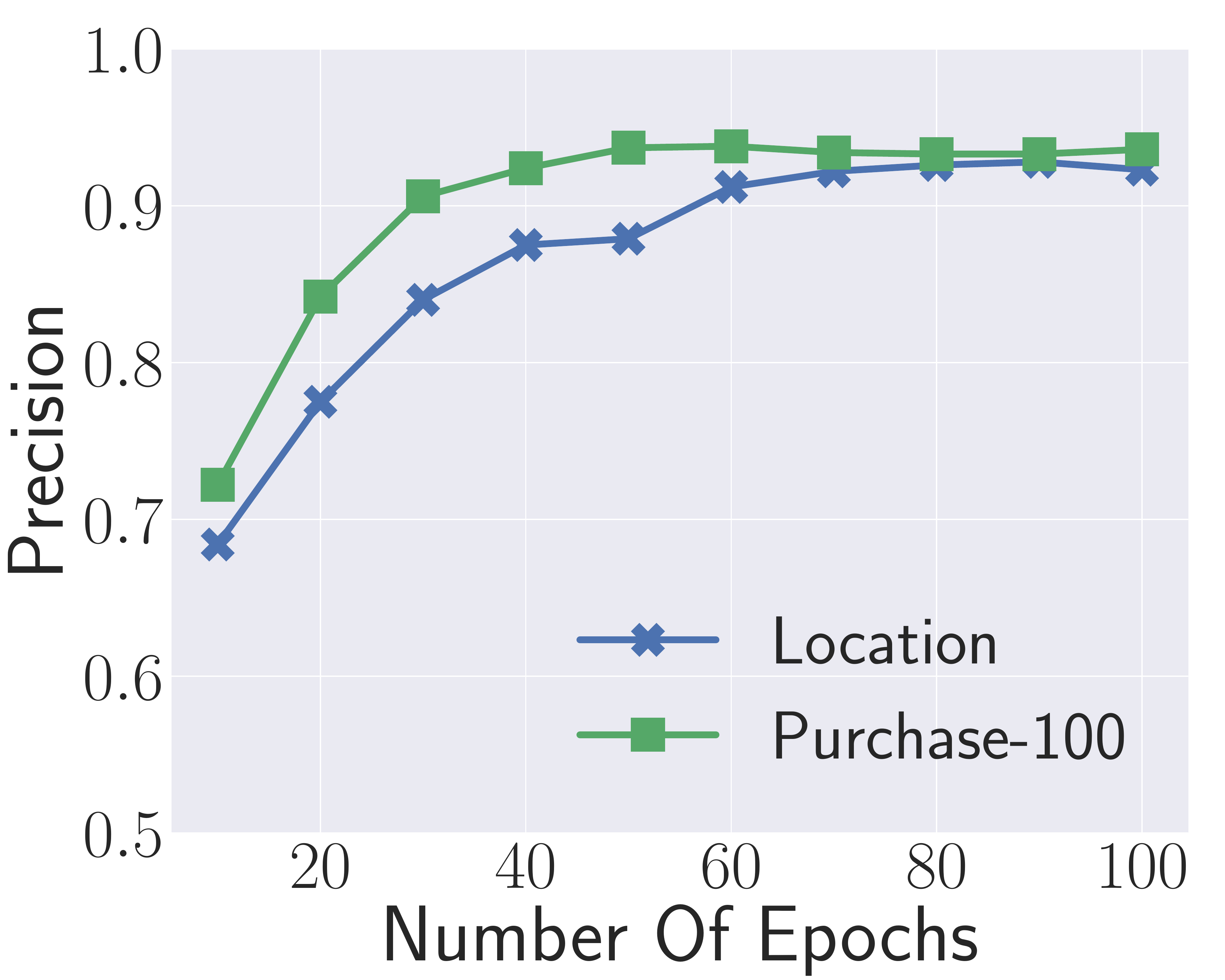}
   \caption{}
   \label{fig:epochsVsPrecision} 
\end{subfigure}
\begin{subfigure}{0.23\textwidth}
   \includegraphics[width=\linewidth]{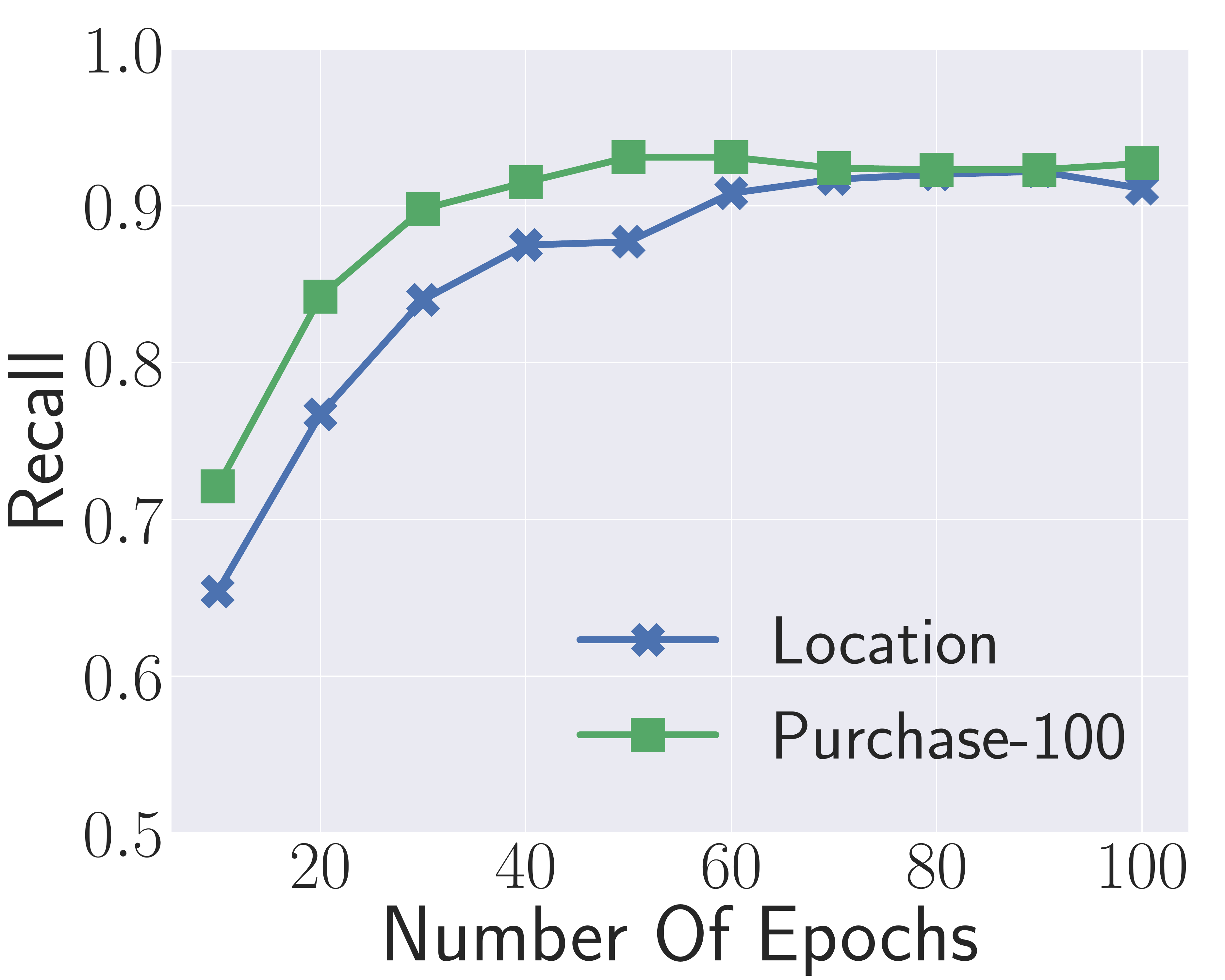}
   \caption{}
   \label{fig:epochsVsRecall}
\end{subfigure}
\caption{The relation between the number of epochs used during the training of the target model (x-axis) and membership inference attack performance (y-axis). (a) precision, (b) recall.}
\label{fig:epochsVsPerformance}
\end{figure}

To further demonstrate the relationship between overfitting and membership inference, we perform another -- more controlled -- experiment on the Location and Purchase-100 datasets.
Concretely, we focus on the number of epochs used in training, larger number leads to higher overfitting.
We vary the number of epochs used from 10 to 100 and report the result in \autoref{fig:epochsVsPerformance}.
As we can see, the attack performance indeed increases with the increase of number of epochs.

Another factor which also affects our attack's performance 
is the number of classes in the dataset.
Both CIFAR-10 and CIFAR-100 are image datasets
with different number of classes (10 vs 100),
it turns out that our membership inference attack on the latter dataset
achieves a 10\% better performance than on the former dataset.
Similar results can be observed from all the Purchase datasets.

\begin{figure}[!ht]
\centering
\begin{subfigure}{0.23\textwidth}
   \includegraphics[width=\linewidth]{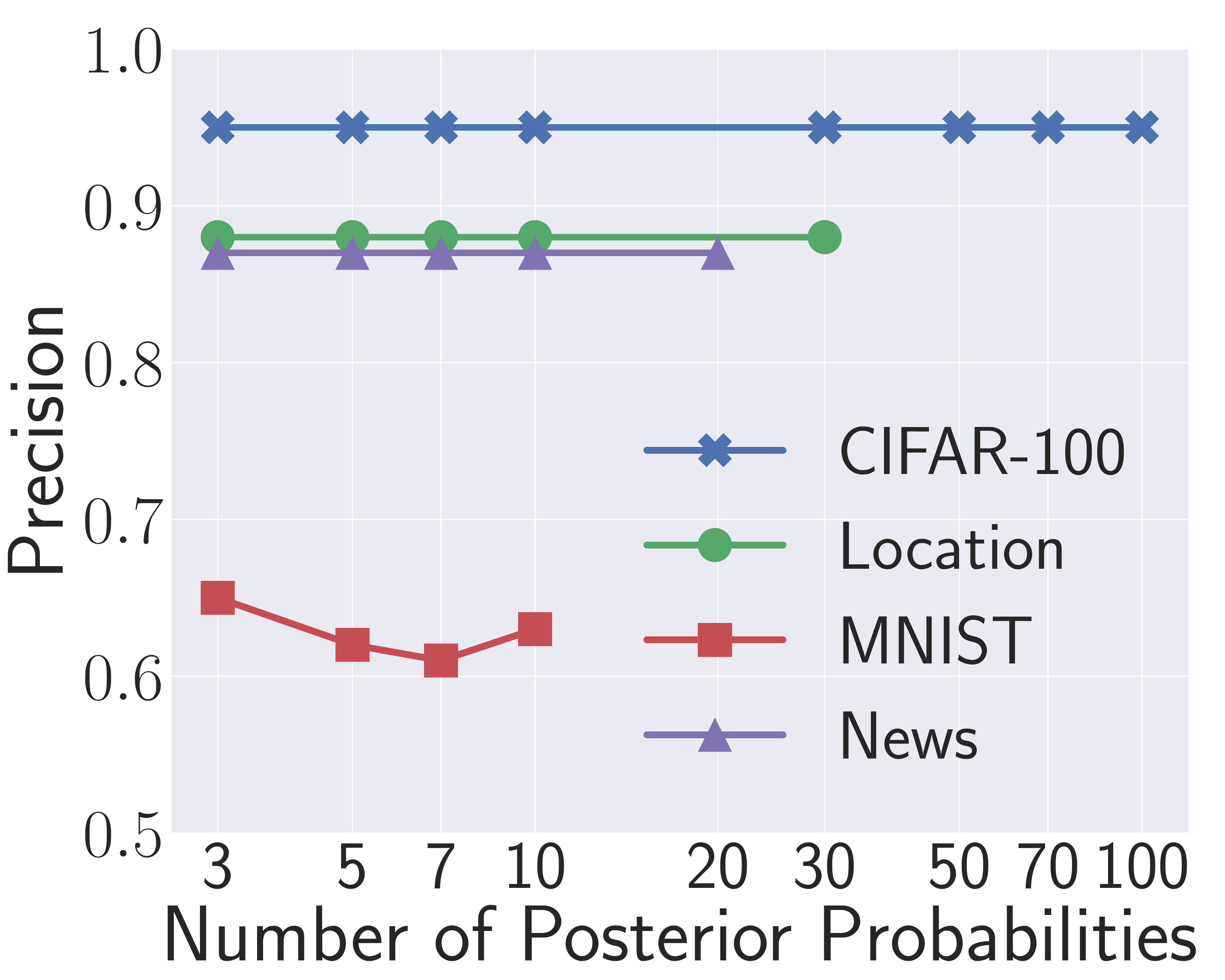}
  \caption{}
\end{subfigure}
\begin{subfigure}{0.23\textwidth}
   \includegraphics[width=\linewidth]{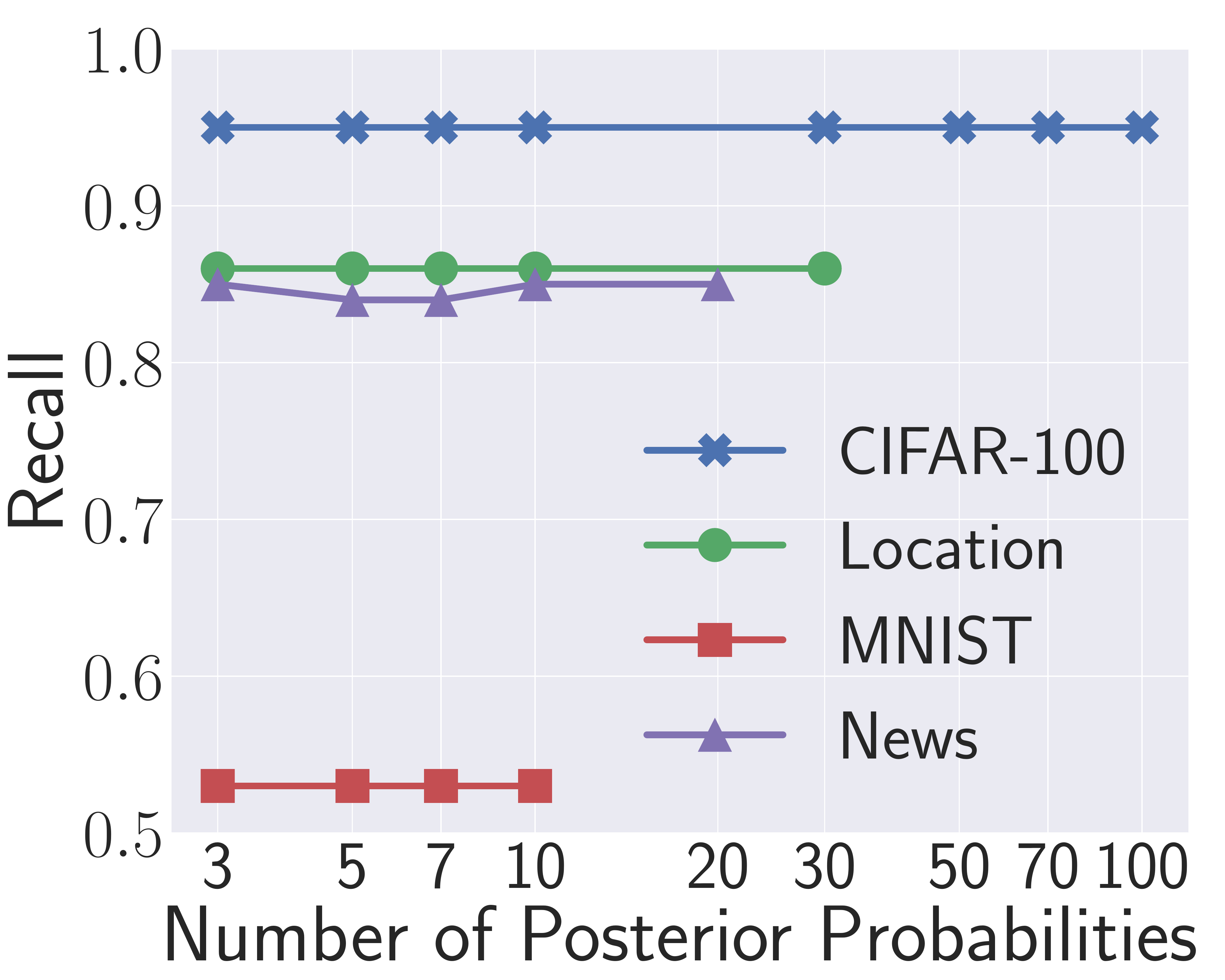}
   \caption{}
\end{subfigure}
\caption{The effect of the number of posterior probabilities (used as features) on the first adversary's performance. (a) precision, (b) recall.}
	\label{fig:featNum} 
\end{figure}

For our attacks, we use only the three highest posterior probabilities (in descending order)
as the features for our attack.
We test the effect of using more posteriors
on the CIFAR-100, Location, MNIST, and News datasets.
The result in \autoref{fig:featNum} shows that this factor does not have a significant effect
on the attack's performance for most of the datasets.
Generally, three posteriors achieves the best performance,
especially on the MNIST dataset.

\begin{figure}[!ht]
\centering
\begin{subfigure}{0.23\textwidth}
   \includegraphics[width=\linewidth]{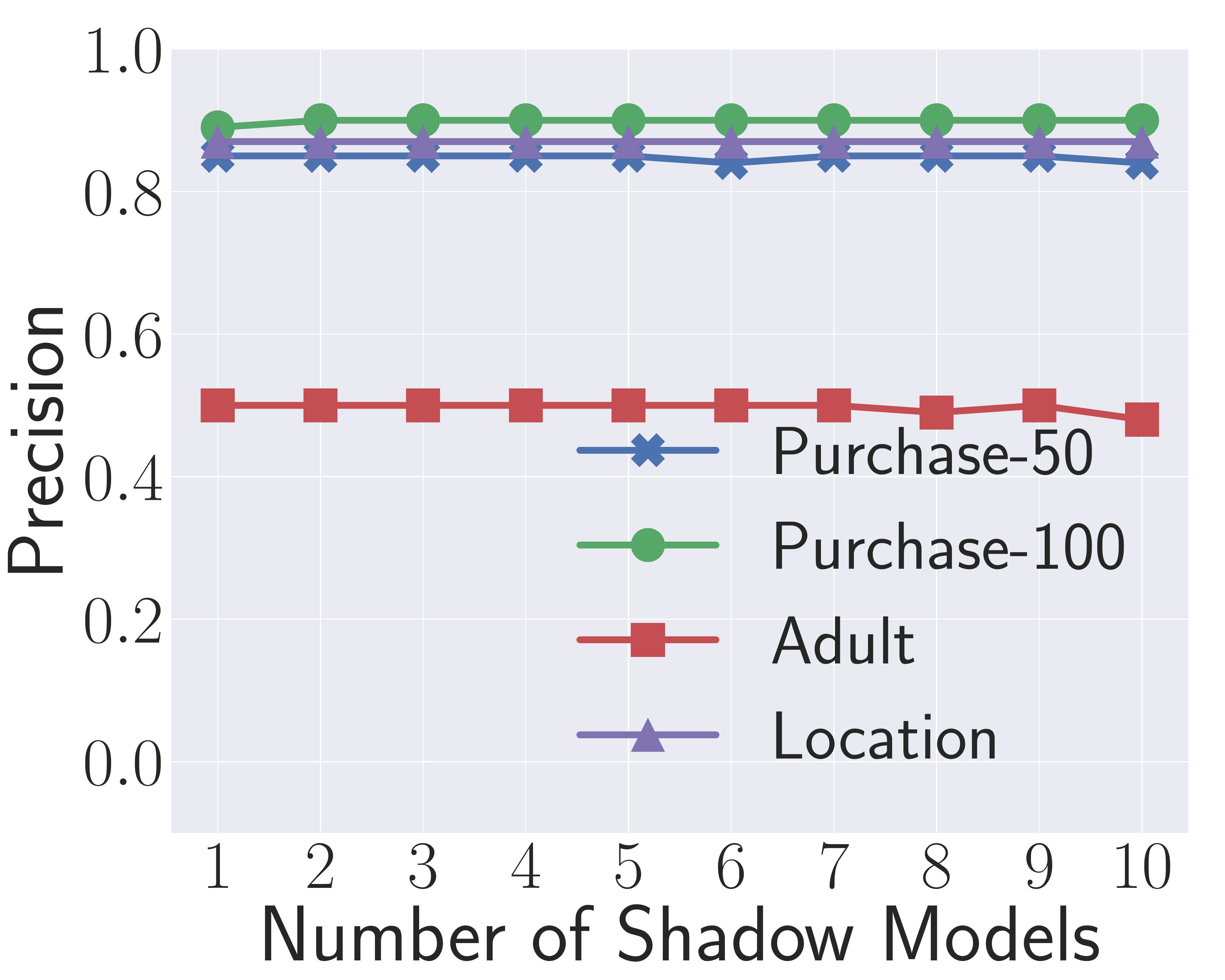}
  \caption{}
	\label{fig:shadowModelsPrec} 
\end{subfigure}
\begin{subfigure}{0.23\textwidth}
   \includegraphics[width=1\linewidth]{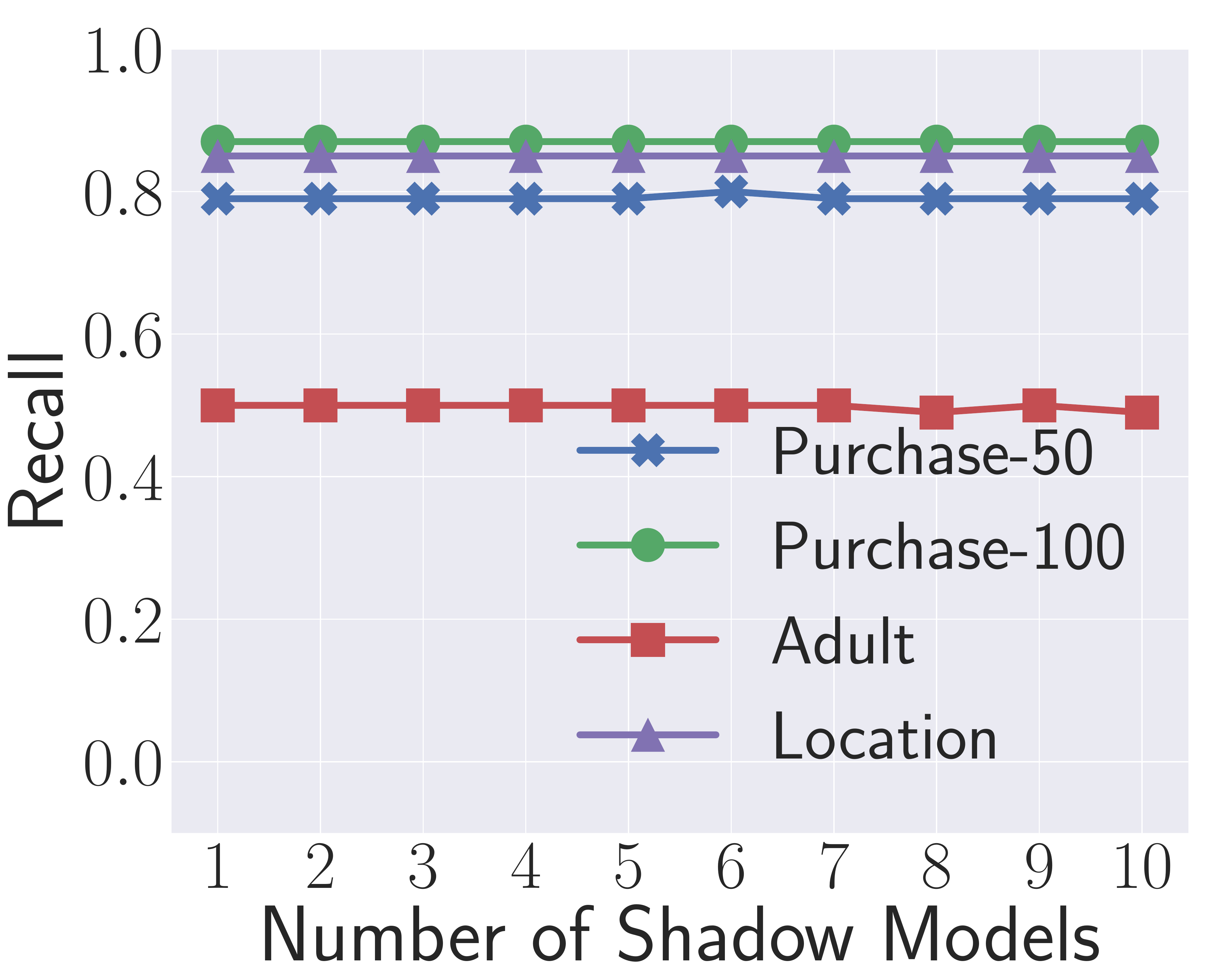}
   \caption{}
	\label{fig:shadowModelsRecall} 
\end{subfigure}
\caption{The effect of the number of shadow models on the first adversary's performance. (a) precision, (b) recall.}
\label{fig:shadowModels} 
\end{figure}

A major difference between our attack and the previous one~\cite{SSSS17}
is the number of shadow models used.
We further study this factor's influence on our own attack's performance.
\autoref{fig:shadowModels} shows the corresponding results on the Purchase-100, 
Purchase-50, Adult, and Location datasets.
By varying the number of shadow models from 1 to 10,
we do not observe a significant performance difference for both precision and recall.
This means increasing the number of shadow models does not improve our attack's performance.

\noindent\textbf{Evaluation on MLaaS.}
All the above experiments are conducted in a local setting.
We further evaluate our attack with a real-world MLaaS.
In particular, we use Google's MLaaS, 
namely Google Cloud Prediction API.\footnote{
Google's Cloud prediction API is deprecated on April 20th, 2018 
(\url{https://cloud.google.com/prediction/}),
the new MLaaS provided by Google is called Cloud Machine Learning Engine.
}
Under this service,
a user can upload her own data and get the black-box ML API trained by Google.
The user can neither choose which classifier to use, nor the corresponding model structure and parameters.
We perform our attack following the same methodology as in \autoref{sec:attack1Meth}.
We construct both target model and shadow model with Google's MLaaS
and build our attack model locally.

We use the Purchase-100 and Location datasets for evaluation
and observe that the attack's performance is even stronger 
than our previous local evaluation.
For the Purchase-100 dataset,
our attack on Google's MLaaS
has a 0.90 precision and a 0.89 recall,
while our local evaluation has a 0.89 precision and a 0.86 recall.
For the Location dataset, 
the precision is 0.89 and the recall is 0.86, 
which is almost similar to our local evaluation (0.88 precision and 0.86 recall).

\subsection{Target Model Structure}
\label{sec:simpTheAssum}
One of the above attack's assumptions is that
the adversary knows the target model's algorithm and hyperparameters
and implements her shadow model in the same way.
Next, we show how to relax this assumption.
We first concentrate on target model's hyperparameters,
then, the type of classifiers it uses.

\noindent\textbf{Hyperparameters.}
We assume that the adversary knows the target model is a neural network,
but does not know the details of the model.
We first train the shadow model with half of the training parameters of the target model.
More precisely, we reduce the batch size,  hidden units, and regularization parameters to half.
On the Purchase-100 dataset,
our attack achieves a 0.86 precision and 0.83 recall,
which is almost the same as the one reported in \autoref{fig:compareWithShokri}.
We also revert the settings to test the case 
when the shadow model has double number of parameters than the target model.
The performance drops a bit to 0.82 precision and 0.80 recall, but it is still quite close to our original attack.
We also perform evaluation over other datasets and observe similar results.
This evaluation shows the flexibility of the membership inference attack:
An adversary with no knowledge about the model's hyperparameters 
can still get good performance.

\begin{figure}[!t]
\centering
\includegraphics[width=0.8\columnwidth]{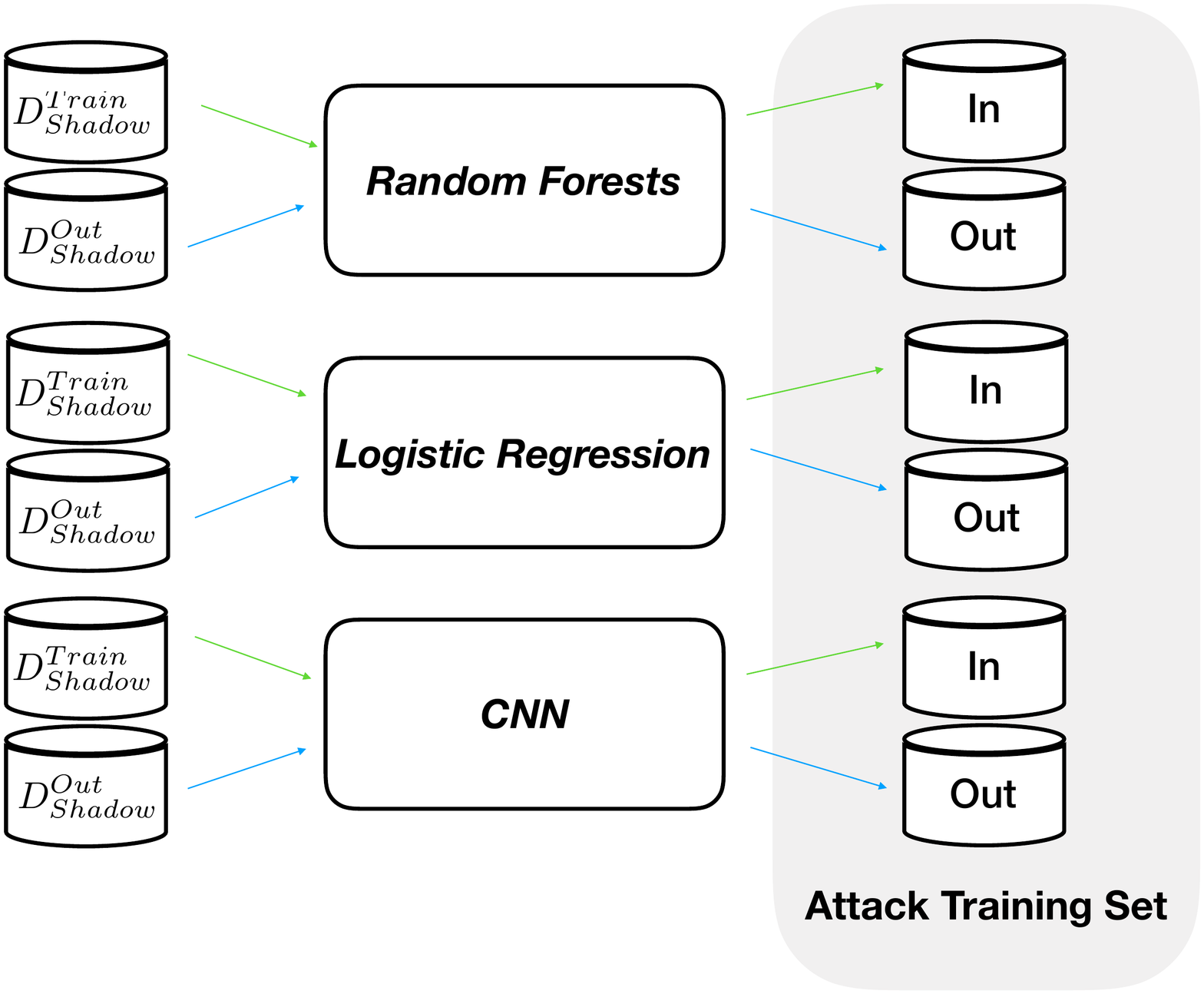}
\caption{The architecture of the combining attack
on generating data for training the attack model.}
\label{fig:noModelData}
\end{figure}

\noindent\textbf{Target Model's Algorithm.}
We further assume that the adversary has no knowledge 
on what classification algorithm is adopted by the target model.
In this setting, our first attempt
is to use any classifier, such as random forests, as the shadow model
and attack the target model that is (very likely to be) different from the shadow model,
such as CNN.
However, the experimental results are not very promising.

To improve the attack with no knowledge of the target model,
we construct a set of ML models,
each with a different classification algorithm,
and combine them together as one shadow model.
Each single ML model is referred to as a \emph{sub-shadow model}.
This is achievable as the types of classifiers are limited.
This attack, also referred to as the \emph{combining attack},
can learn the behavior of the different classifiers 
and therefore can attack an unknown target model 
based on the assumption that there is a sub-shadow model which is trained with the same classifier as the target model.

Concretely, we use the same methodology as in \autoref{sec:attack1Meth}
to train multiple sub-shadow models as illustrated in \autoref{fig:noModelData}, 
with each sub-shadow model being a different classifier.
The data each sub-shadow model is trained on is the same.
All the generated features by all sub-shadow models are stacked together,
i.e., the attack model~$\attack$ is trained with a larger dataset.
In this new dataset, each data point in~$\shadowData$ 
is represented multiple times with respect to different sub-shadow models' outputs.

We run a local experiment on the Purchase-100 dataset
to evaluate this attack.
Three popular ML classifiers, i.e., multilayer perceptron, 
random forests (with 1,000 tree), and logistic regression
are adopted as sub-shadow models.
The target model for the Purchase-100 dataset
is a multilayer perceptron.
For a more complete comparison,
we further build another two target models 
that are based on random forests and logistic regression, respectively,
and use the same algorithm to build a single shadow model as in \autoref{sec:attack1Meth}.
\autoref{tab:multiModels} depicts the result.
As we can see, our combining attack has a similar performance
when target model is multilayer perceptron
and logistic regression.
Meanwhile, the attack's performance is relatively worse
is when the target model is random forests.

In conclusion,
we show that our combining attack can free the attacker from knowing the target model,
which further enlarges the scope of membership inference attack.

\begin{table}[!t]
\centering
\scalebox{1}
{
\begin{tabular}{lcccc} 
 \toprule
 \multirow{2}{*}{Classifier}  &  \multicolumn{2}{c}{With target model structure}& 
 \multicolumn{2}{c}{Combining attack} \\ 
\cmidrule(lr){2-3}\cmidrule(lr){4-5}
 & Precision &Recall
 & Precision &Recall    \\\midrule
Multilayer perceptron &0.86&0.86&0.88&0.85\\
Logistic regression&0.90&0.88&0.90&0.88\\
Random forests &1.0&1.0&0.94&0.93\\\bottomrule
\end{tabular}
}
\caption{Comparison of the combining attack 
and the original attack by the first adversary proposed in \autoref{sec:attack1Meth}.}
\label{tab:multiModels}
\end{table}
\begin{figure*}[!ht]
\centering
\begin{subfigure}{0.49\textwidth}
   \includegraphics[width=1\linewidth]{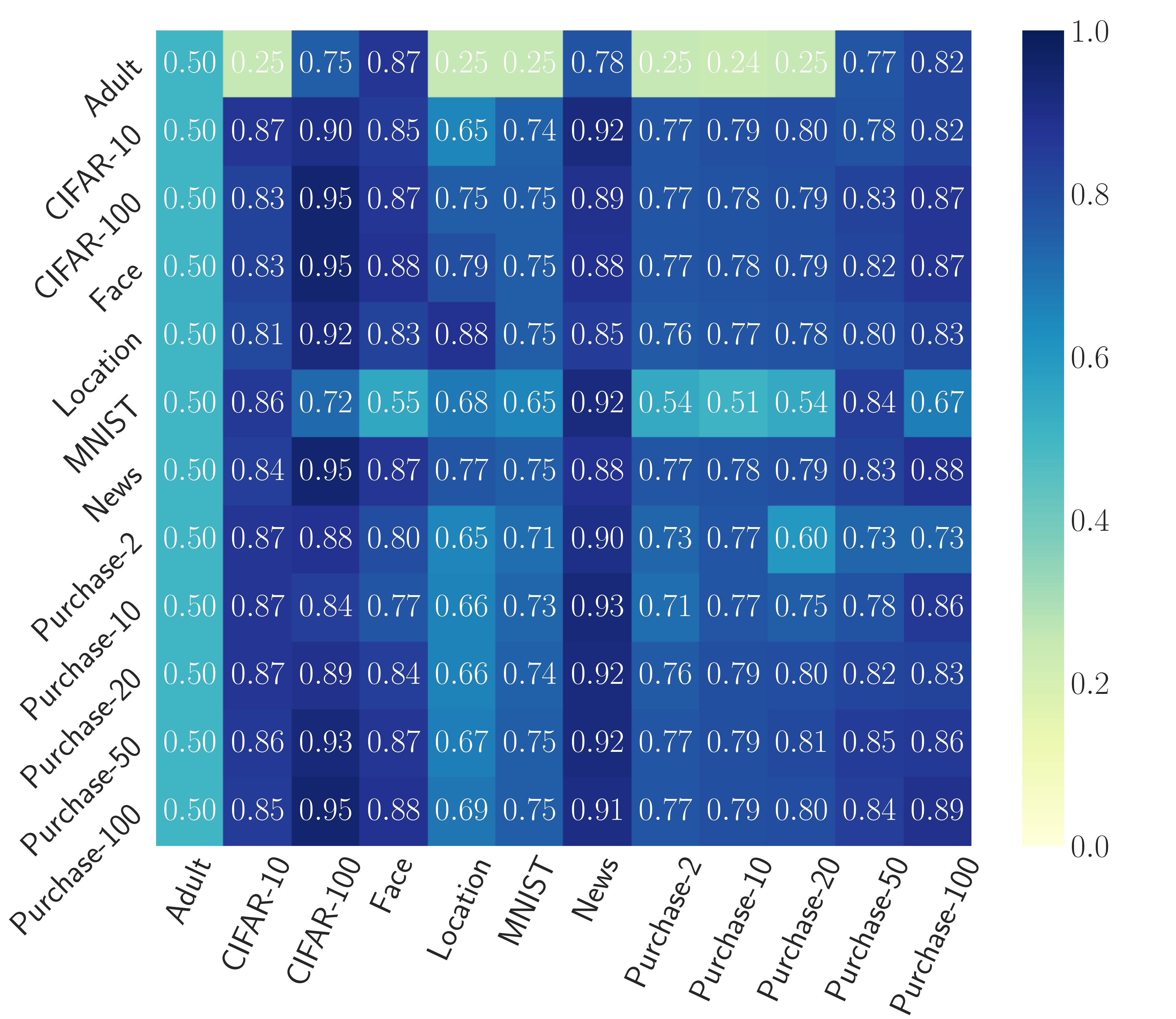}
   \caption{Precision}
   \label{fig:datasetTransPrec} 
\end{subfigure}
\begin{subfigure}{0.49\textwidth}
   \includegraphics[width=1\linewidth]{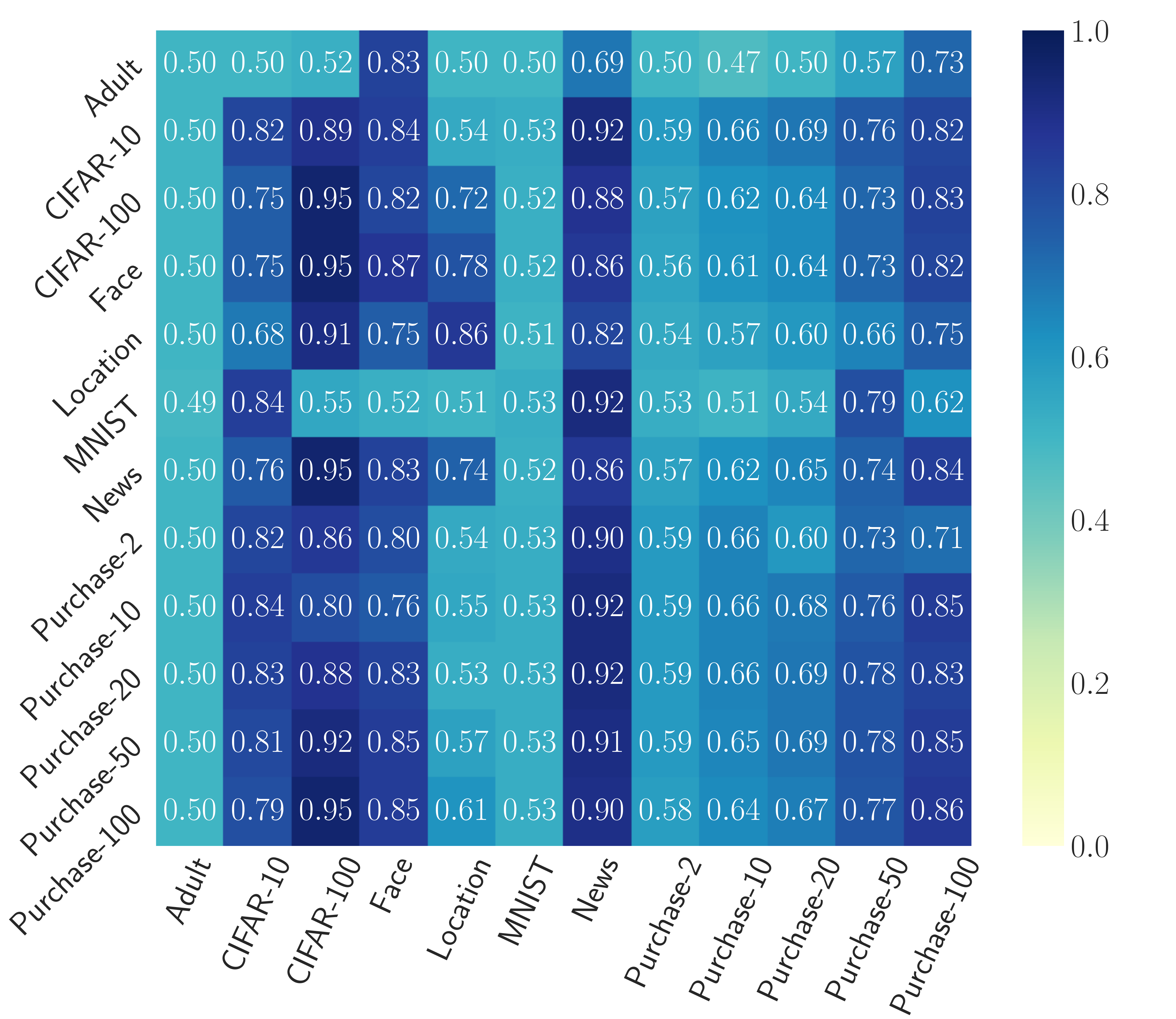}
   \caption{Recall}
   \label{fig:datasetTransRecall}
\end{subfigure}
\caption{
The performance of our data transferring attack.
The x-axis represents the dataset being attacked, i.e., the dataset the target model is trained on.
The y-axis represents the dataset used for training the shadow model.
}
\label{fig:datasetTrans}
\end{figure*}

\section{Towards Data Independent Membership Inference Attacks (Adversary 2)}
\label{sec:attack2}

In this section,
we relax the assumption on the adversary having a dataset
that comes from the same distribution
as the target model's dataset.

We start by explaining the threat model,
then describe the adversary's attack methodology.
In the end, we present a comprehensive experimental evaluation.

\subsection{Threat Model}

Different from the threat model in \autoref{sec:attack1},
we remove the assumption that the adversary has a dataset $\shadowData$ coming from the same distribution as the training data for the target model.
This largely reduces the attack capabilities of the adversary.
For this scenario,
Shokri et al.~\cite{SSSS17} propose to query the target model multiple times
to generate synthetic data to train the shadow model.
However, this approach can only be applied 
when the dataset is assembled 
with binary features.\footnote{We confirm this with the authors.}
In contrast, our approach can be applied to attack ML models trained on any kind of data.

\subsection{Methodology}
\label{sec:attack2Meth}

The strategy of the second adversary
is very similar to the one of the first adversary.
The only difference is that 
the second adversary utilizes an existing dataset 
that comes from a different distribution than the target model's training data
to train her shadow model.
We refer to this attack as the \emph{data transferring attack}.

The shadow model here is not to mimic the target model's behavior,
but only to summarize the membership status of a data point in the training set of a machine learning model.
As only the three - or two in case of binary datasets - largest posteriors are used for the attack model,
we can also neglect the effect brought by datasets with different number of classes.

\subsection{Evaluation}

\noindent\textbf{Experimental Setup.}
We use the same attack model and shadow model setup 
as presented in \autoref{sec:attack1},
such as data splitting strategy and the types of ML models used.
We perform the data transferring attack over all datasets.
For evaluation metric, we again use precision and recall.

\begin{figure}[!t]
\centering
\begin{subfigure}{0.23\textwidth}
\centering
   \includegraphics[width=\linewidth]{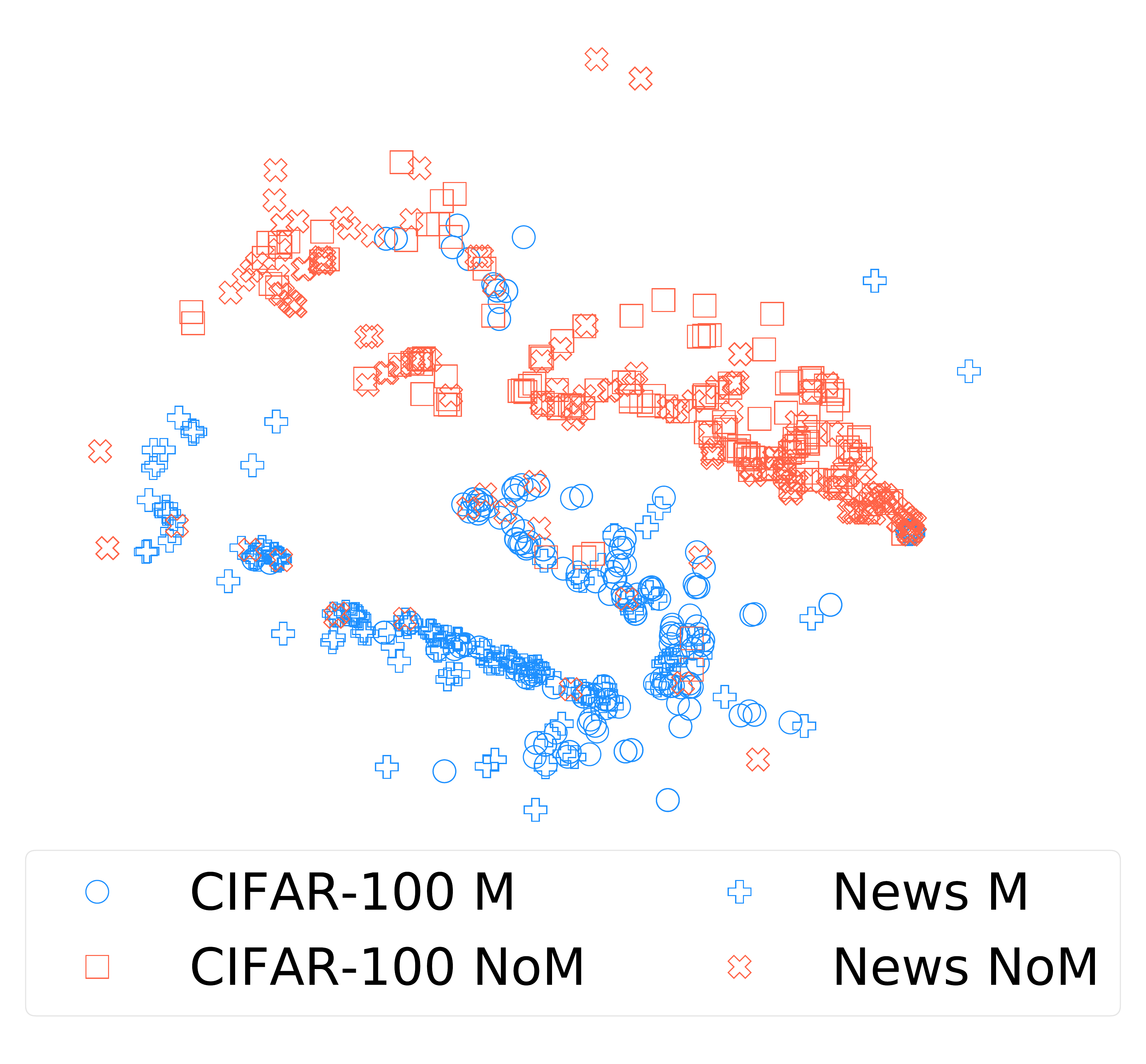}
   \caption{}
   \label{fig:whyAttack2Work} 
\end{subfigure}
\begin{subfigure}{0.23\textwidth}
\centering
   \includegraphics[width=\linewidth]{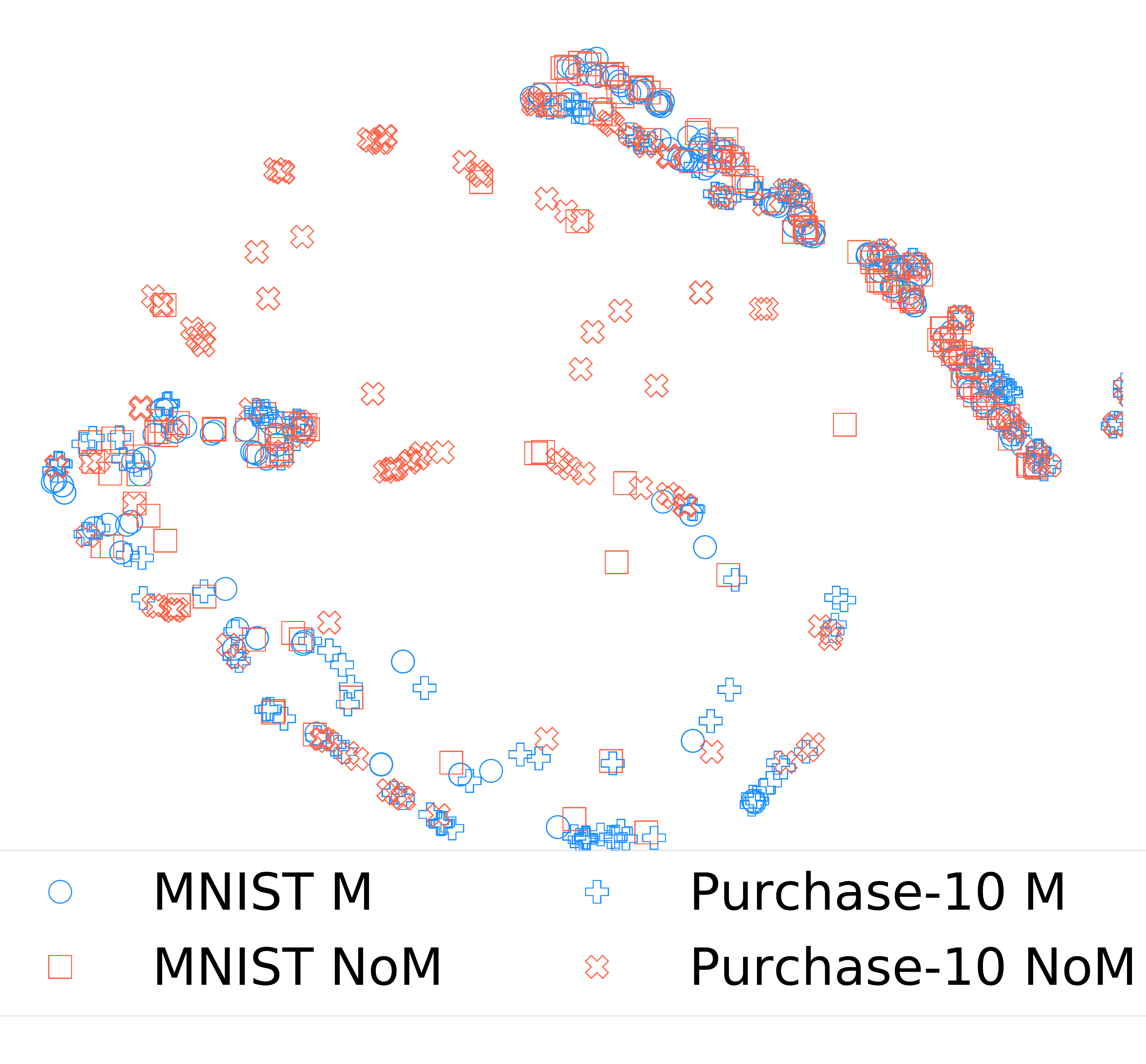}
   \caption{}
   \label{fig:whyAttack2NotWork}
\end{subfigure}
\caption{The top three posteriors of member and non-member data points (a random sample)
projected into a 2D space using t-Distributed Stochastic Neighbor Embedding (t-SNE).
(a) CIFAR-100 and News, (b) MNIST and Purchase-10.
M means member and NoM means non-member.}
\label{fig:whyAttack2}
\end{figure}

\noindent\textbf{Results.}
\autoref{fig:datasetTrans} depicts the data transferring attack's performance.
The x-axis 
represents the dataset being attacked, i.e., the dataset the target model is trained on,
and the y-axis represents
the dataset used for training the shadow model.
Compared to the first adversary
the attack results of which are listed at the diagonal of \autoref{fig:datasetTrans}, 
the second adversary in multiple cases obtains similar performances.
For instance, using the Face dataset to attack the CIFAR-100 dataset 
results in 0.95 for both precision and recall,
while the corresponding results for the first adversary are also 0.95 for both metrics.
In several cases, we even observe a performance improvement over the first adversary.
For instance,
using the Purchase-10 dataset to attack the News dataset achieves a 0.93 precision and 0.92 recall,
while the first adversary has a 0.88 precision and 0.86 recall.
More interestingly,
in many cases, datasets from different domains can effectively attack each other,
e.g., the News dataset and the CIFAR-100 dataset.

For the first adversary,
we relax the assumption on shadow model design.
This relaxation also applies for the second adversary,
as the shadow model and target model
are trained with different datasets.
For instance, the Purchase-20 dataset is trained with a multilayer perceptron
while the CIFAR-100 dataset is trained with a CNN.

One of the major advantages of our data transferring attack
lies in its applicability.
The synthetic data generation strategy by Shokri et al.~\cite{SSSS17} 
cannot be applied to dataset of any kind, but those with binary features.
Even for dataset of binary features,
a single synthetic data point requires 156 queries~\cite{SSSS17}
to the target model.
Given the large dataset quantity needed for ML models
and MLaaS's pay-per-query business model,
this is very costly.
Moreover, sending a large amount of queries to an MLaaS API
would alert the server,
which may not even allow the adversary to finish her synthetic data generation process.
Meanwhile, our data transferring attack does not have any of the above constraints.

\noindent\textbf{Reasoning.}
After demonstrating the strong performance of our data transferring attack, we now seek to understand the reason behind.
To this end, we pick the highest three posteriors (similar to our attack) of member and non-member data points with respect to their target ML models of all datasets, and embed these posteriors into a 2D space using t-Distributed Stochastic Neighbor Embedding (t-SNE).
We show in \autoref{fig:whyAttack2Work} the result for two datasets (of different types) between which our transferring attack is effective.
As we can see, the member and non-member points of these datasets are tightly clustered together and follow a common decision boundary,
thus, the attack model trained on one dataset
can effectively infer the membership status of points in the other dataset.
Meanwhile, \autoref{fig:whyAttack2NotWork} shows the results
for two datasets between which our transferring attack is not effective.
As depicted, there are no clear clusters for members and non-member data points.

\subsection{Evaluation On MLaaS}
\label{sec:mlaas2}

We also evaluate our data transferring attack on Google's MLaaS.
Concretely, we use a shadow model trained on the Location dataset
to attack a target model trained on the Purchase-100 dataset.
Both models are trained with Google's MLaaS.
Experimental results show that we achieve a 0.8 precision and 0.78 recall.
By further flipping the shadow and target model, i.e., Purchase-100 dataset attacking Location dataset,
the membership inference result is still very strong with a 0.87 precision and a 0.82 recall.
This shows that our data transferring attack is not only effective in the local setting
but also in the real-world MLaaS setting.

\section{Model and Data Independent Membership Inference Attack without Training (Adversary 3)}
\label{sec:attack3}

In this section,
we present our third adversary,
who does not need to train any shadow model and does not assume knowledge of model or data distribution.
We start with the threat model description.
Then, we list the attack methodology.
In the end, we present the evaluation results.

\subsection{Threat Model}

We relax the assumption 
that the adversary needs to train any shadow model to perform her attack.
All she could rely on is the target model's output posteriors $\model(\point)$ 
after querying her target data point $\point$.
Note that Yeom et al.~\cite{YGFJ18} propose a similar attack,
however, their membership inference attack
requires the adversary to know the target data point's class label
which is hard to obtain in some cases, such as in biomedical settings~\cite{BHZLEB18}.
Therefore, our threat model covers a broader range of scenarios.

\subsection{Methodology}
\label{sec:attack3Meth}

The attack model for the third adversary
is implemented as an unsupervised binary classification.
Concretely, the adversary first obtains $\model(\point)$.
Then, she extracts the highest posterior 
and compares whether this maximum is above a certain threshold.
If the answer is yes, then she predicts the data point is in the training set of the target model and vice versa.
The reason we pick maximum as the feature
follows the reasoning that an ML model is more confident, 
i.e., one posterior is much higher than others,
when facing a data point that it was trained on.
In another words,
the maximal posterior of a member data point is much higher than the one of a non-member data point.

\begin{figure}[!t]
\centering
\includegraphics[width=1\linewidth]{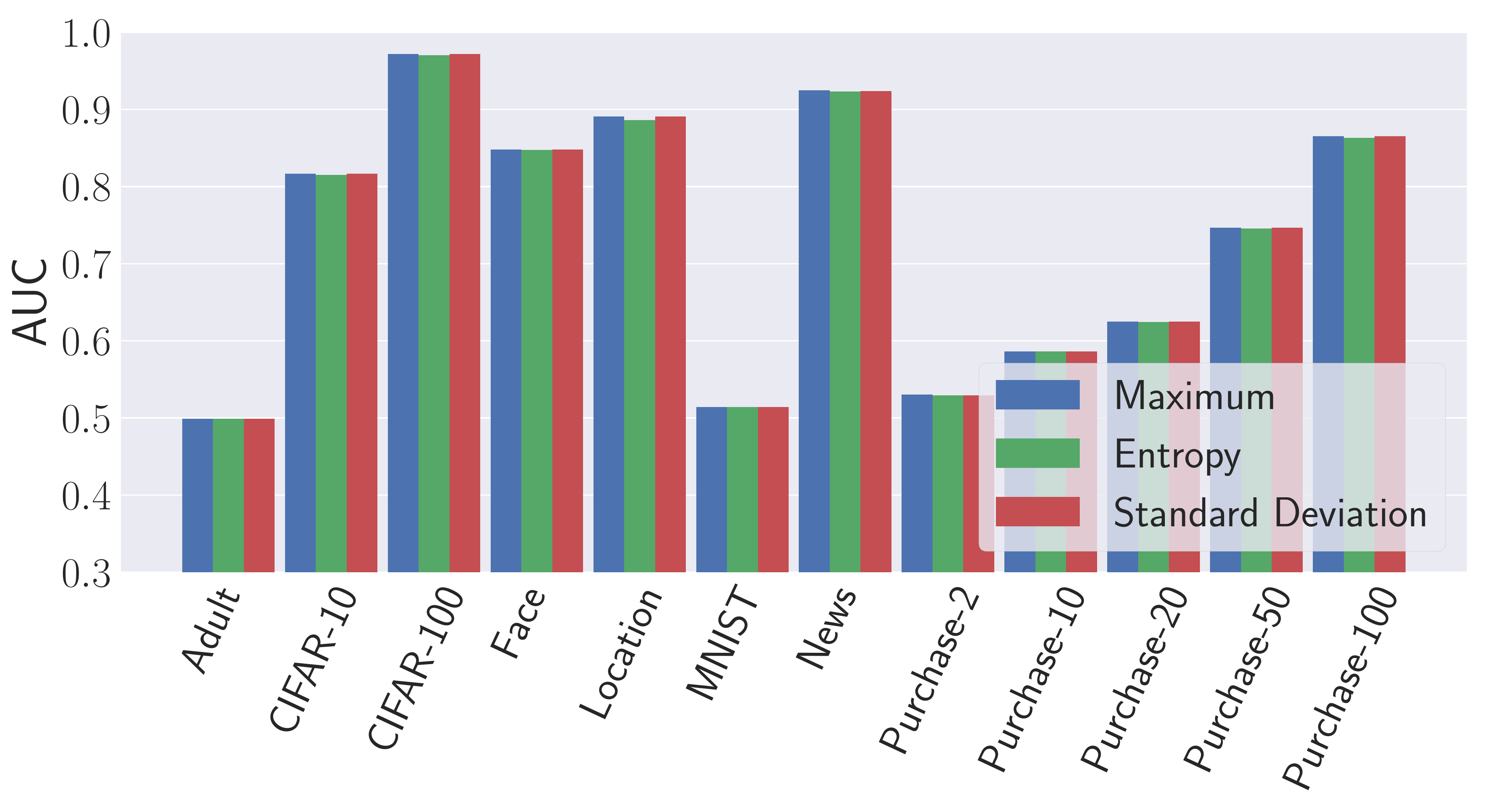}
\caption{The AUC values for three different statistical measures over all datasets. 
We include the results for the Adult and News datasets as 
AUC is independent of a concrete detection threshold.
}
\label{fig:whyMaxAttk3}
\end{figure}

\noindent\textbf{Threshold Choosing.}
The attacker can pick the threshold for membership inference 
depending on her requirements,
as in many machine learning applications~\cite{ZHSMVB17,BHPZ17}.
For instance, if she concentrates more on inference precision (recall),
then she can pick a relatively high (low) threshold.

Nevertheless, we provide a general method for choosing a threshold.
Concretely,
we generate a sample of random points in the feature 
space of the target data point.
For image datasets including CIFAR-10, CIFAR-100, MNIST, and Face, 
we generate random images, where the value of each pixel is drawn from a uniform distribution.
For datasets with binary features including Location and Purchase datasets, 
we generate 0 and 1 for each feature according to an unbiased coin flip.
For Adult and News, as the bounds for features are not clear,
our method cannot apply.
One way to tackle this is to collect News articles or people's records (with the same features as in the Adult dataset) from the Internet as the ``random'' points.
We leave this for future work.
Next, we query these random points to the target model to get the corresponding maximal posteriors.
We hypothesize that these points act as the non-member points.
Thus, top $t$ percentile of these random points' maximal posteriors can serve as a good threshold.
Below, we show empirically that there exists a choice of $t$ percentile 
that works well and generalizes across all the dataset and therefore can be used to automatically determine the detection threshold.

\subsection{Evaluation}

\noindent\textbf{Experimental Setup.}
We evaluate the third adversary
over all datasets except News and Adult.
Note that we do not need to split the dataset
as this adversary does not train any shadow model.
Instead, we split each dataset by half, and use one part to train the target model
and the other part is left out as non-members.

\begin{figure}[!t]
\centering
\begin{subfigure}{0.47\textwidth}
   \includegraphics[width=\linewidth]{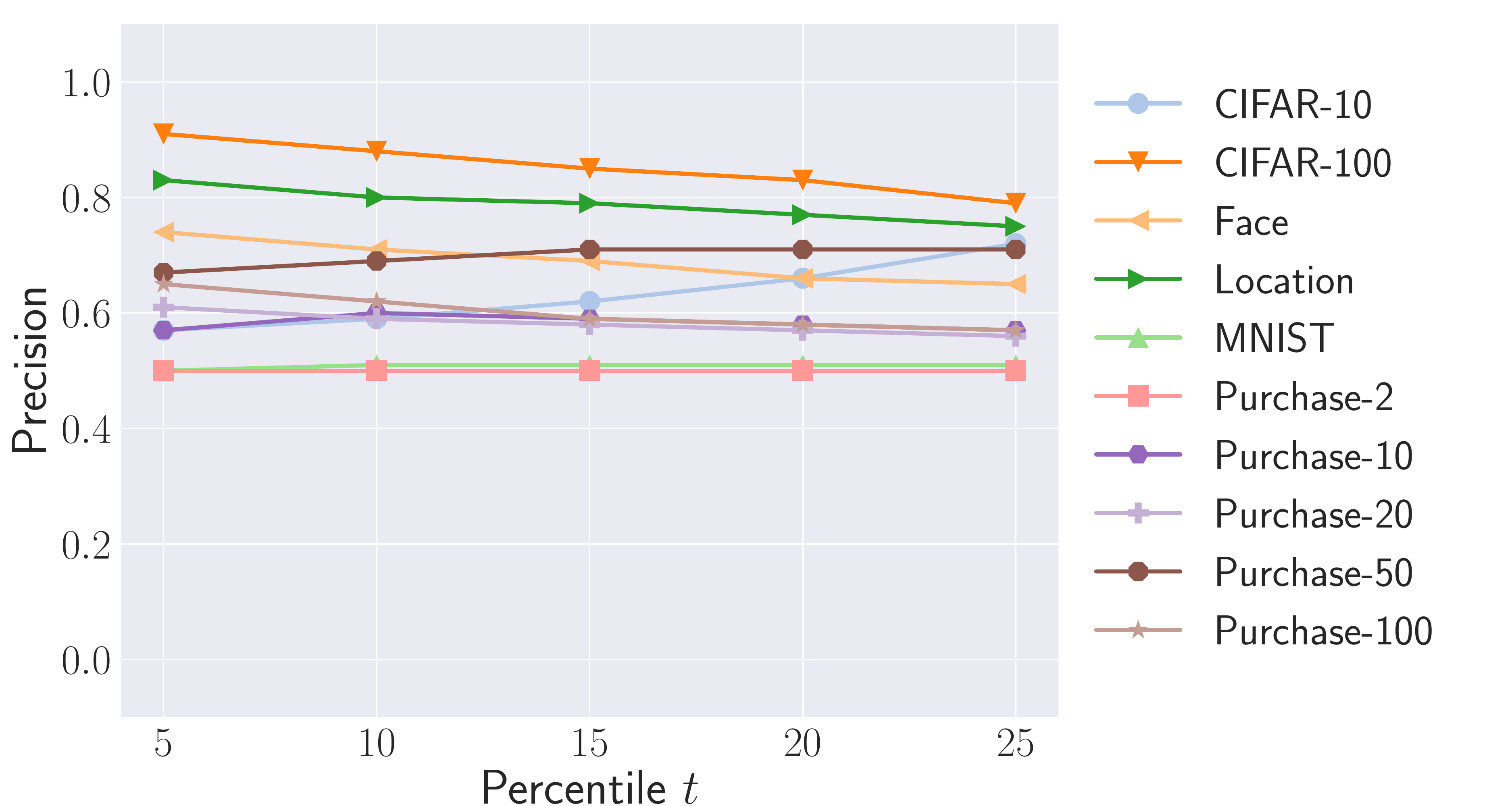}
   \caption{}
   \label{fig:precentilePrec} 
\end{subfigure}
\begin{subfigure}{0.47\textwidth}
   \includegraphics[width=\linewidth]{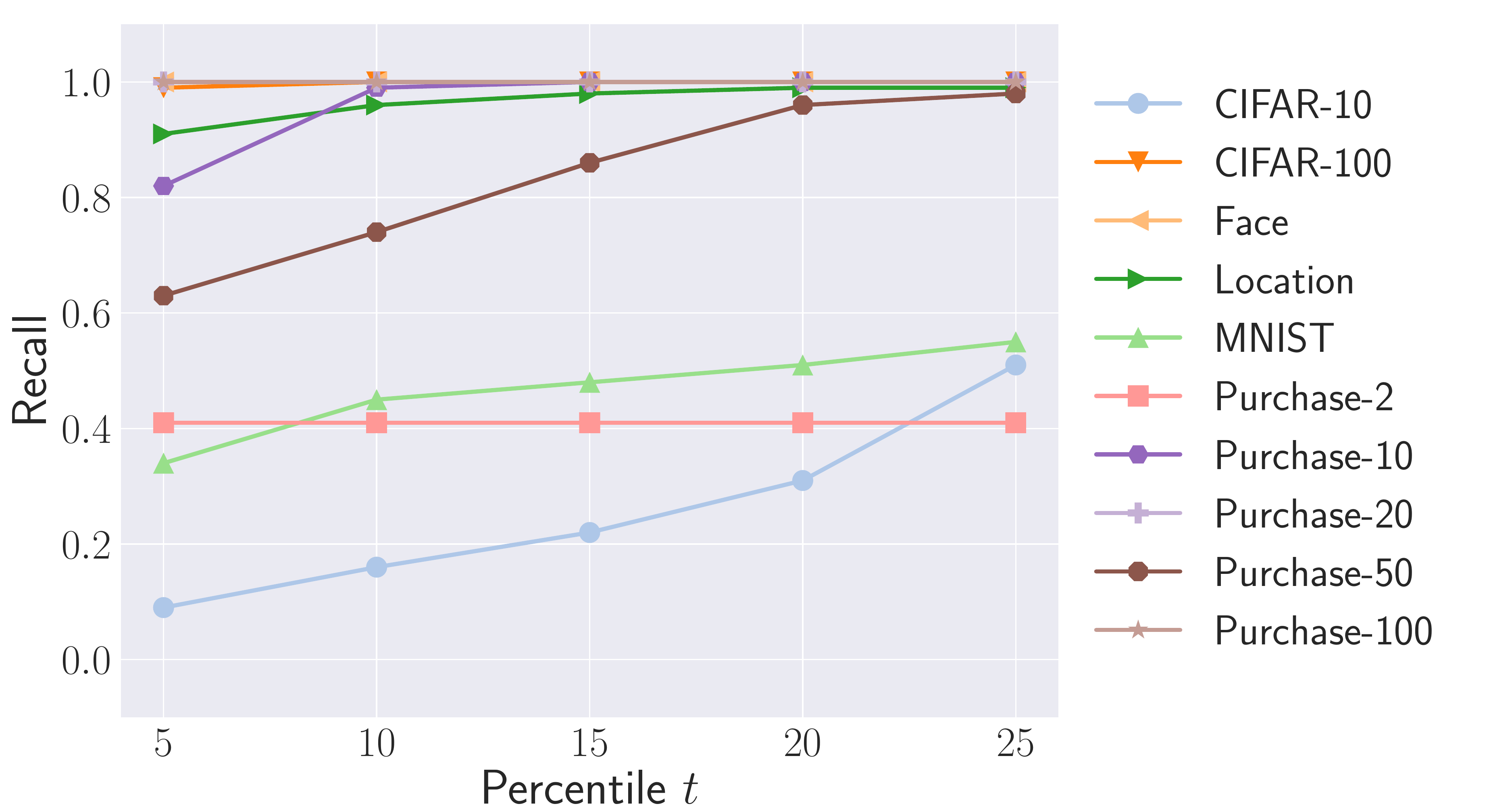}
   \caption{}
   \label{fig:precentileRecall}
\end{subfigure}
\caption{The relation between the percentile of the maximal posterior, i.e., threshold, (x-axis) and the third adversary's performance (y-axis). (a) precision, (b) recall.}
\label{fig:thresholdVsPrecentile}
\end{figure}

\noindent\textbf{Results.}
We first evaluate the effectiveness of maximal posterior
on differentiating member and non-member points without setting a threshold.
To this end, we adopt the AUC (area under the ROC curve) value,
which reports the relation between true positive rate and false negative rate 
over multiple thresholds,
as the evaluation metric~\cite{FLJLPR14,BHPZ17,PTC18,PZ172,ZHRLPB18}.
Besides maximal posterior probability,
we further test the effect of using other statistical metrics, including standard deviation
and entropy.
In particular, the entropy of posteriors is defined as 
$-\sum_{p_i\in \outputvec } p_i \log p_i$,
where $p_i$ denotes the posterior for the $i$-th class.
\autoref{fig:whyMaxAttk3} shows that
maximal posterior achieves very high performance:
In multiple datasets, we obtain AUC values above 0.8.
Meanwhile, the AUC score is almost the same for all the three measures.
This indicates that standard deviation and entropy can also be used as features for the attack.

\begin{figure}[!t]
\centering
\begin{subfigure}{0.23\textwidth}
   \includegraphics[width=1\linewidth]{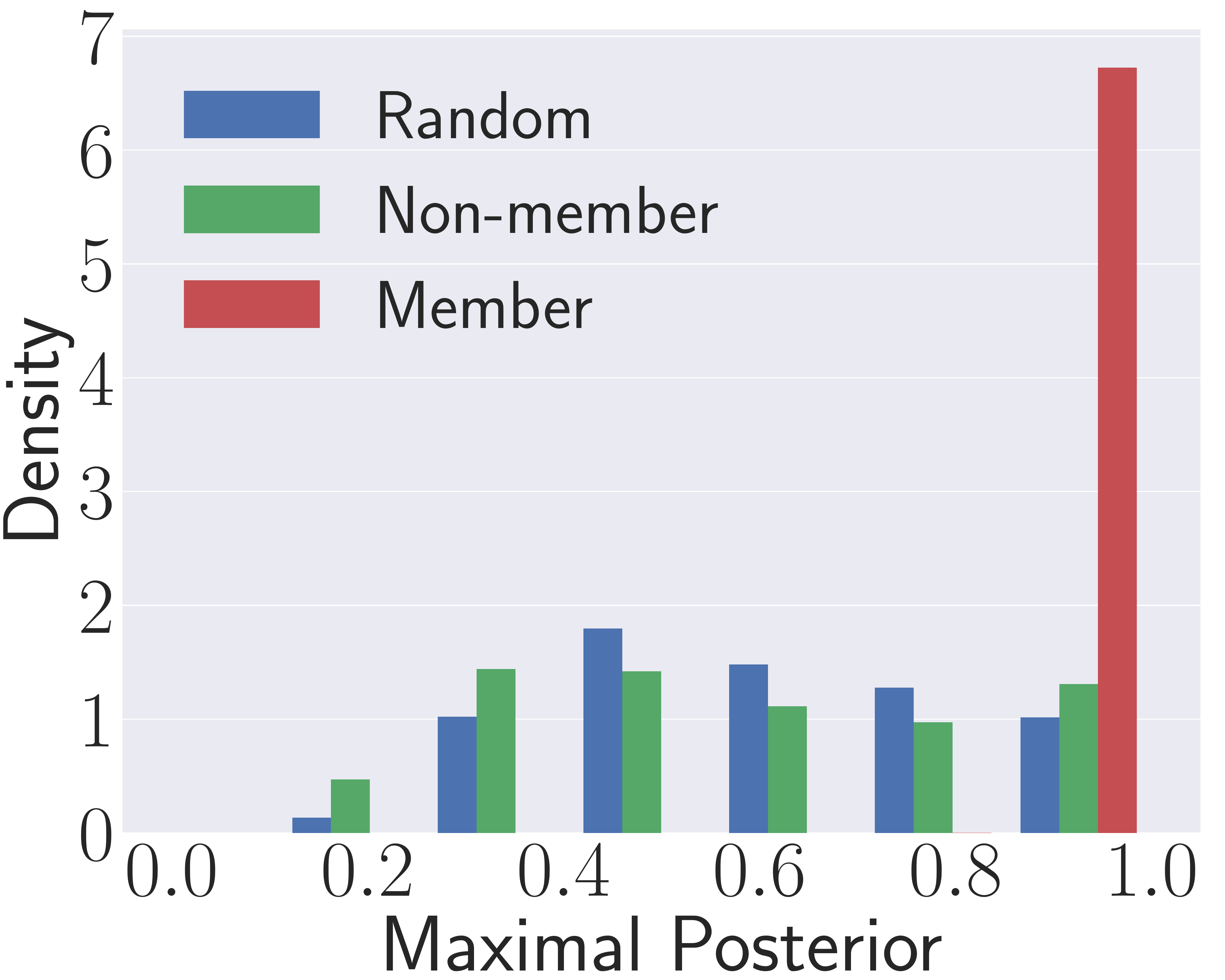}
   \caption{CIFAR-100}
   \label{fig:cifar100Distribution} 
\end{subfigure}
\begin{subfigure}{0.23\textwidth}
   \includegraphics[width=\linewidth]{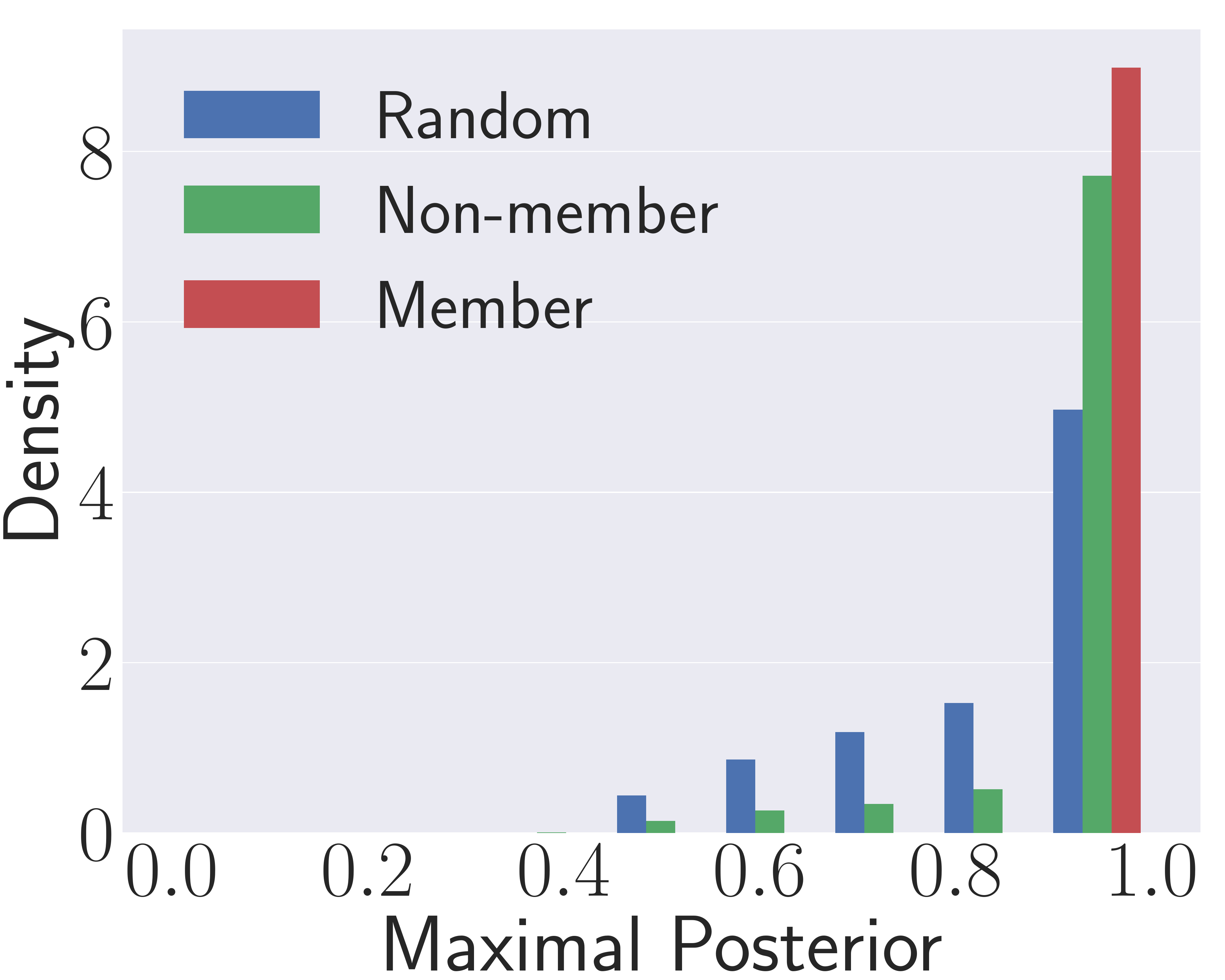}
   \caption{Purchase-10}
   \label{fig:purchase10Distribution}
\end{subfigure}
\caption{The distribution of the maximal posterior of the ML model for the member, non-member, and random points for two datasets: (a) CIFAR-100, (b) News.
The y-axis shows the density.}
\label{fig:attack3Distribution}
\end{figure}

Next, we evaluate our concrete prediction following our threshold-choosing method.
We generate 1,000 random data points for each dataset
and experiment multiple thresholds with respect to the top $t$ percentile.  
\autoref{fig:thresholdVsPrecentile} shows the results.
As we can see, setting $t$ to $10$ achieves a good performance (both precision and recall)
for most of the datasets, such as CIFAR-100.
\autoref{fig:cifar100Distribution} further shows the maximal posterior distribution
of member, non-member, and random points for CIFAR-100.
As the figure shows, our random points' maximal posteriors behave similarly to the distribution of the non-member points' which leads to the strong membership inference.
On the other hand, our attack does not perform well on some datasets, 
such as Purchase-10,
the corresponding maximal posteriors of which are shown in \autoref{fig:purchase10Distribution}.

We also experiment with picking a fixed threshold for membership inference, 
e.g., maximal posterior above 50\%.
However, the evaluation shows that there is no single number that can achieve good performance
for all the datasets.
Thus, we conclude that our threshold-choosing method is suitable.

\begin{figure}[!t]
\centering
\begin{subfigure}[b]{0.47\textwidth}
   \includegraphics[width=1\linewidth]{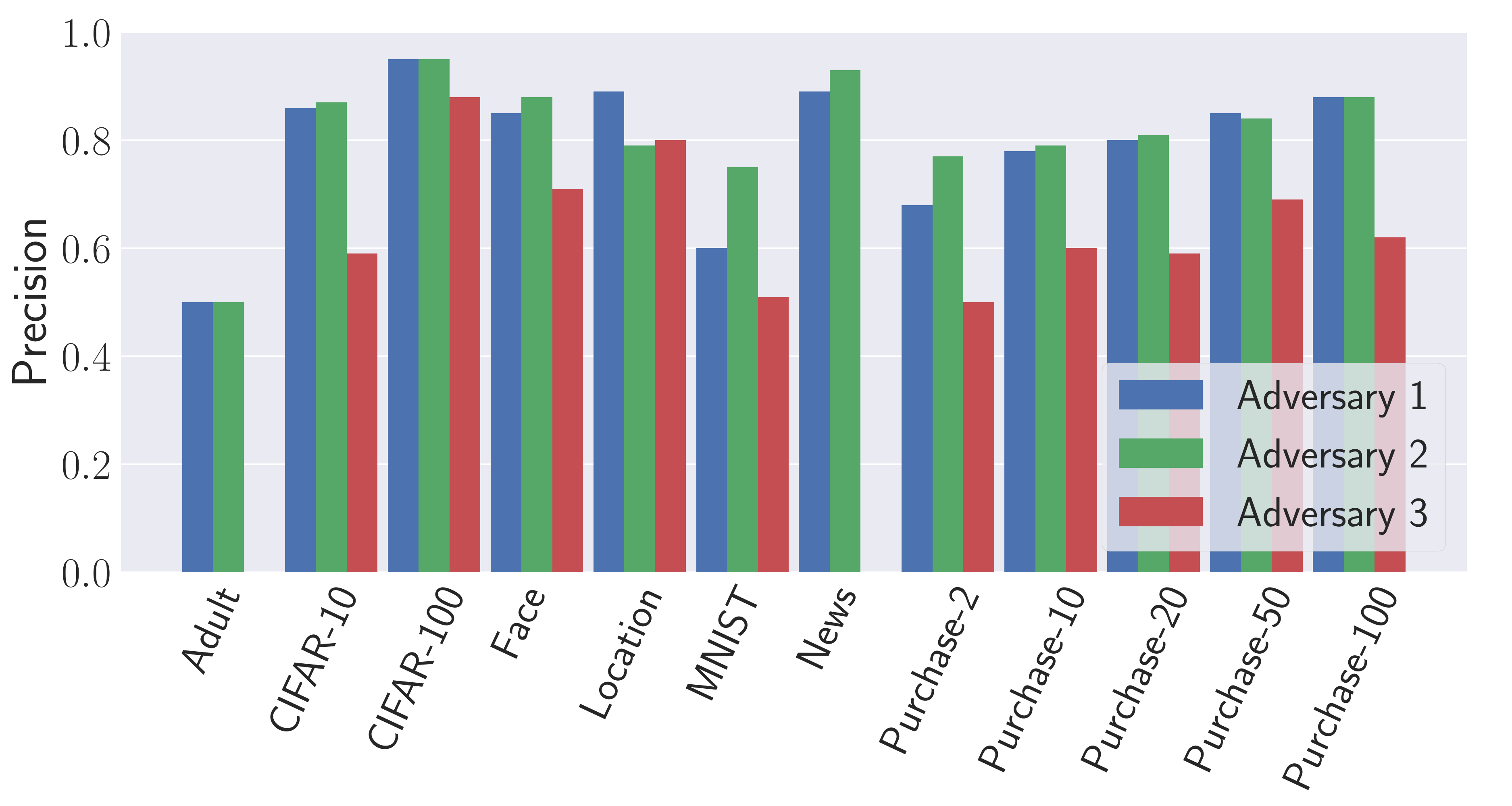}
   \caption{}
   \label{fig:allAttkPrec}
\end{subfigure}
\begin{subfigure}[b]{0.47\textwidth}
   \includegraphics[width=1\linewidth]{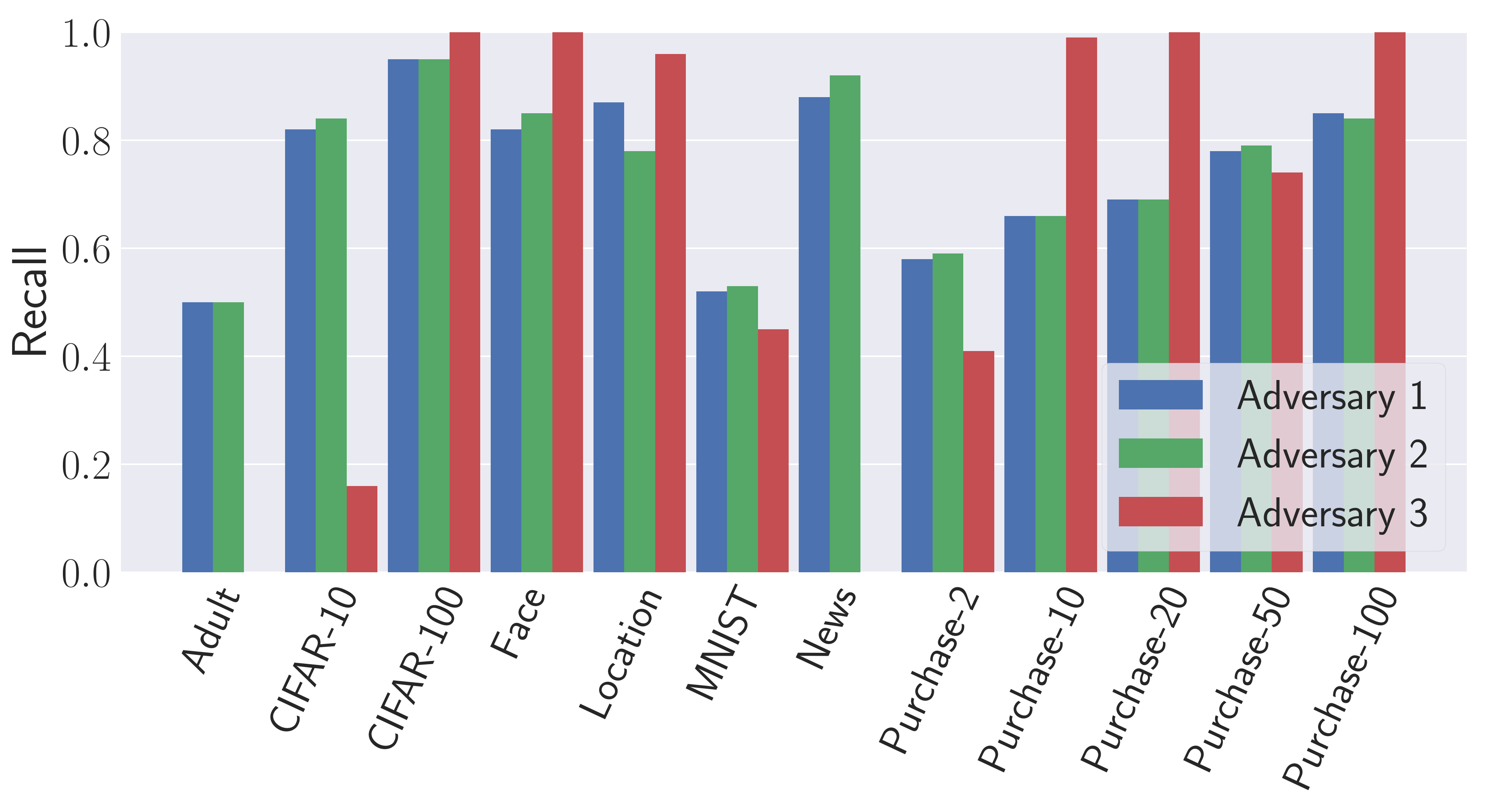}
   \caption{}
   \label{fig:allAttkRecall}
\end{subfigure}
\caption{Comparison of the three different adversaries' performance. (a) precision, (b) recall.}
\label{fig:compareAllAttk}
\end{figure}

\subsection{Comparison of the Three Attacks}

\autoref{fig:compareAllAttk} compares the performance, i.e., precision and recall, of the three attacks.\footnote{All the comparisons are done on the same dataset setting.}
In particular, we show the best performance for our data transferring attack (adversary 2).
As we can see, our first two adversaries achieve very similar performance for most of the datasets.
On the other hand, 
the performance of our third adversary with the minimal assumptions is only slightly worse (especially for precision).
These results clearly demonstrate membership inference attacks
are very broadly applicable,
thereby the corresponding risks are much more severe than previously shown.

\section{Defense}
\label{sec:defense}

In this section,
we propose two defense techniques aiming at mitigating membership privacy risks.
The effectiveness of our membership inference attacks
is mainly due to the overfitting nature of ML models.
Therefore, our defense techniques are designed
to increase ML models' generalizability, i.e., prevent them from being overfitted.

Our first technique is dropout
which is designed for neural network-based classifiers.\footnote{Note that for different kinds of classifiers, dropout can be replaced by other regulation techniques, such as the $L_2$-norm, 
which is also shown to be effective against the membership inference attack \cite{SSSS17}.}
Our second technique is model stacking.
This mechanism is suitable for all ML models, independent of the classifier used to build them.

\begin{figure*}[!h]
\centering
\begin{subfigure}[b]{0.47\textwidth}
   \includegraphics[width=1\linewidth]{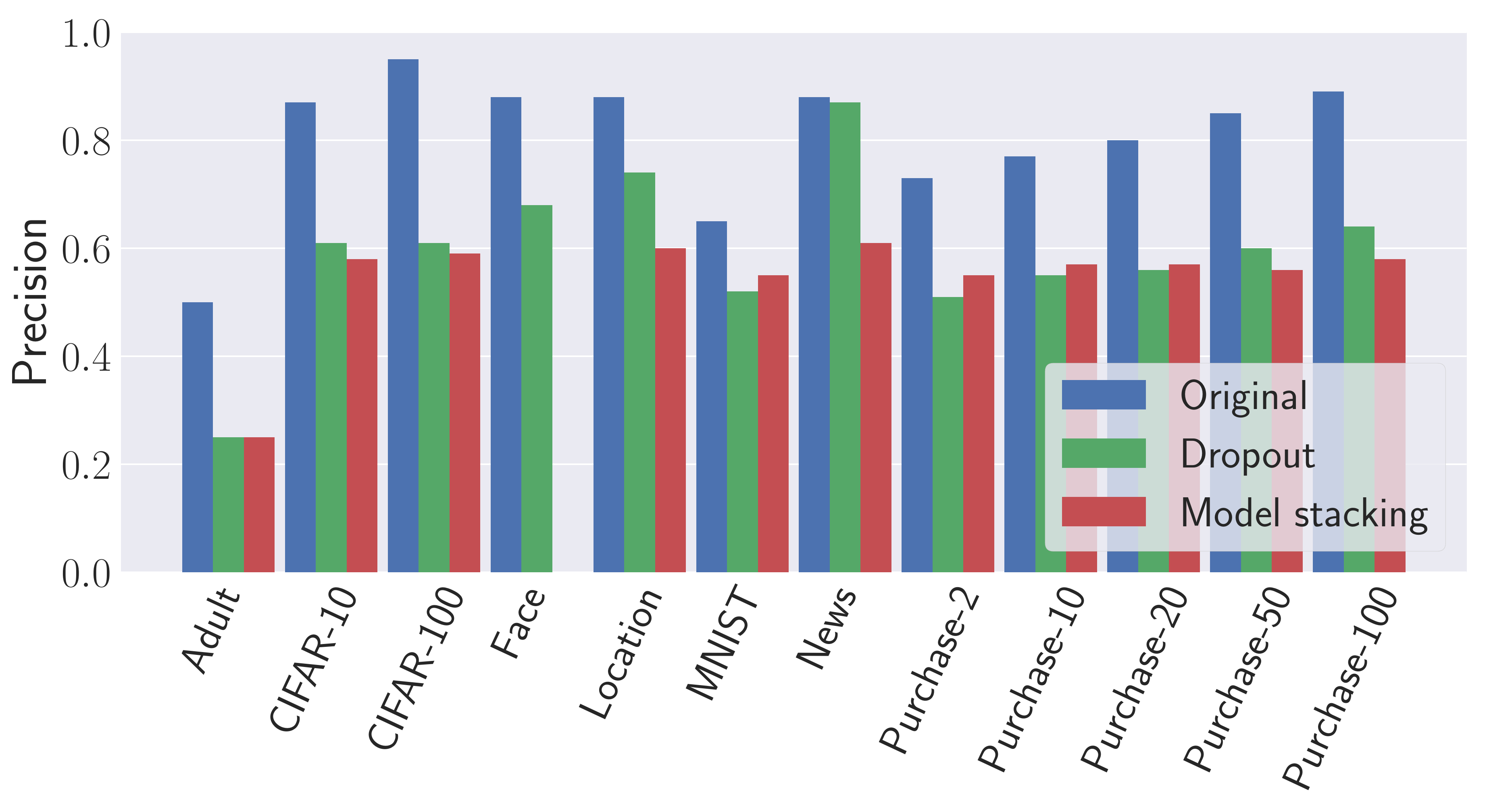}
   \caption{}
   \label{fig:defPres}
\end{subfigure}
\begin{subfigure}[b]{0.47\textwidth}
   \includegraphics[width=1\linewidth]{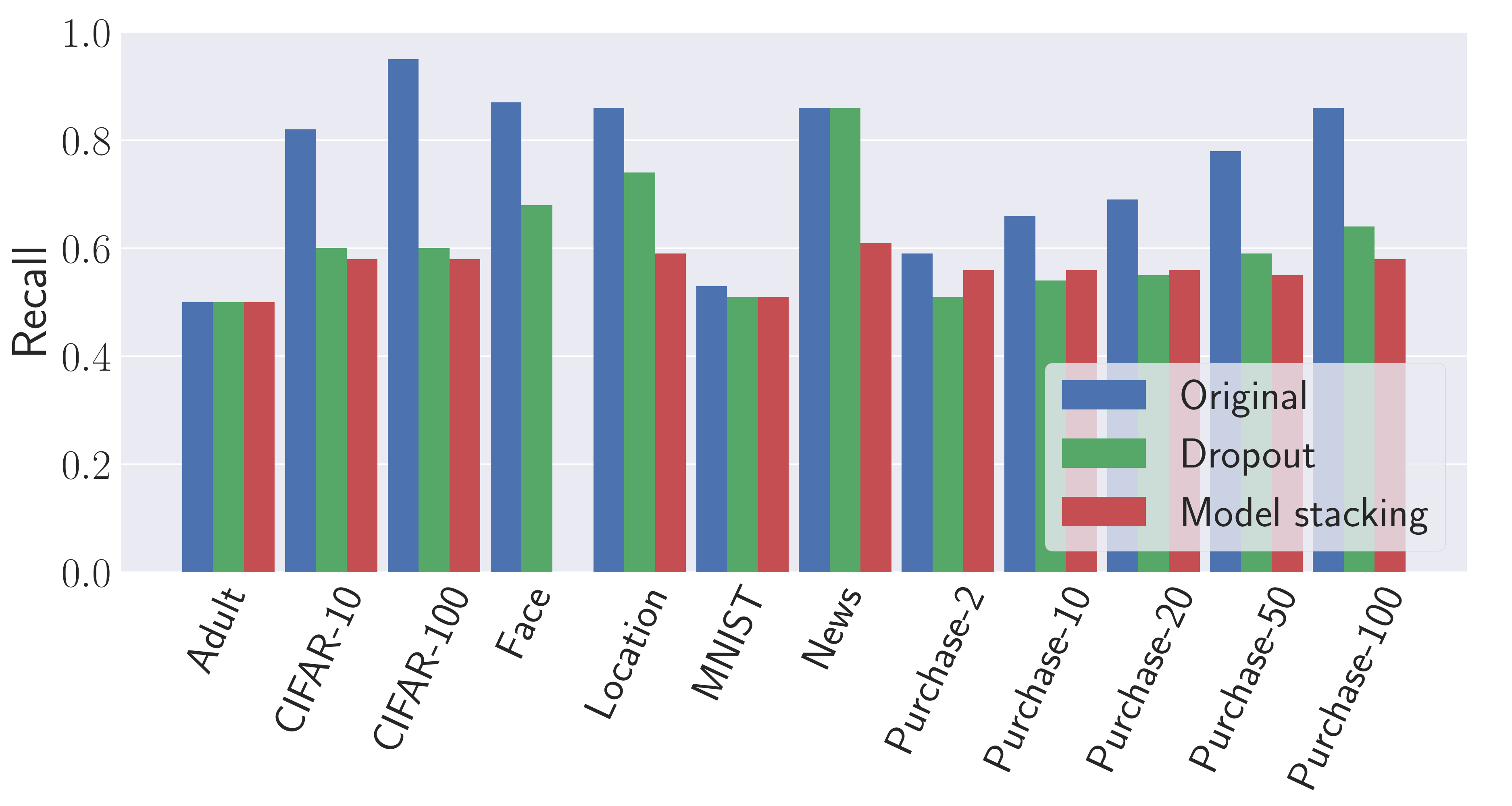}
   \caption{}
   \label{fig:defRecall}
\end{subfigure}
\caption{Comparison of the first adversary's performance 
under both of the defense mechanisms. (a) precision, (b) recall.}
\label{fig:defComp}
\end{figure*}

\begin{figure*}[!t]
\centering
\begin{subfigure}[b]{0.47\textwidth}
   \includegraphics[width=1\linewidth]{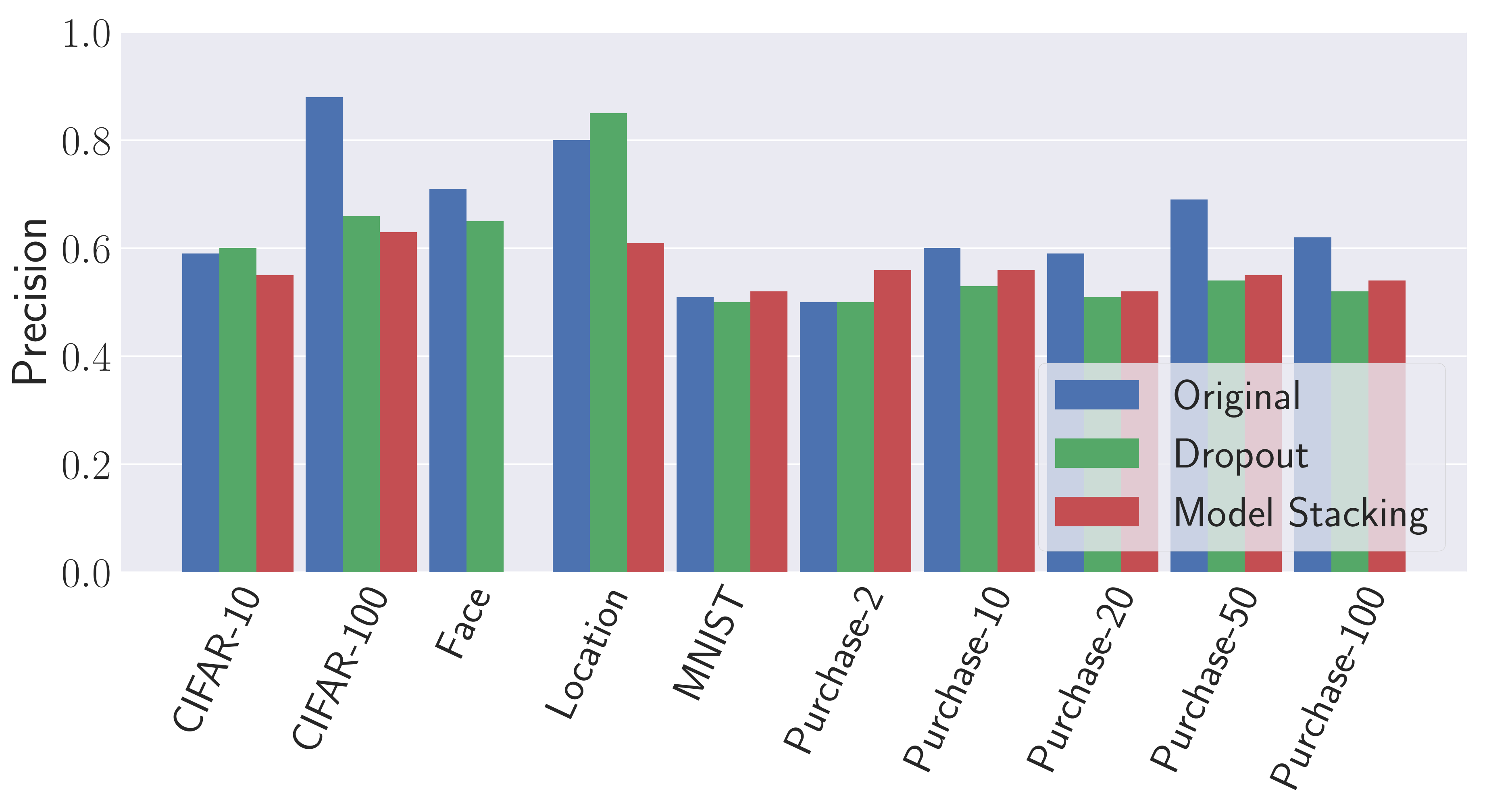}
   \caption{}
   \label{fig:attk3defPres}
\end{subfigure}
\begin{subfigure}[b]{0.47\textwidth}
   \includegraphics[width=1\linewidth]{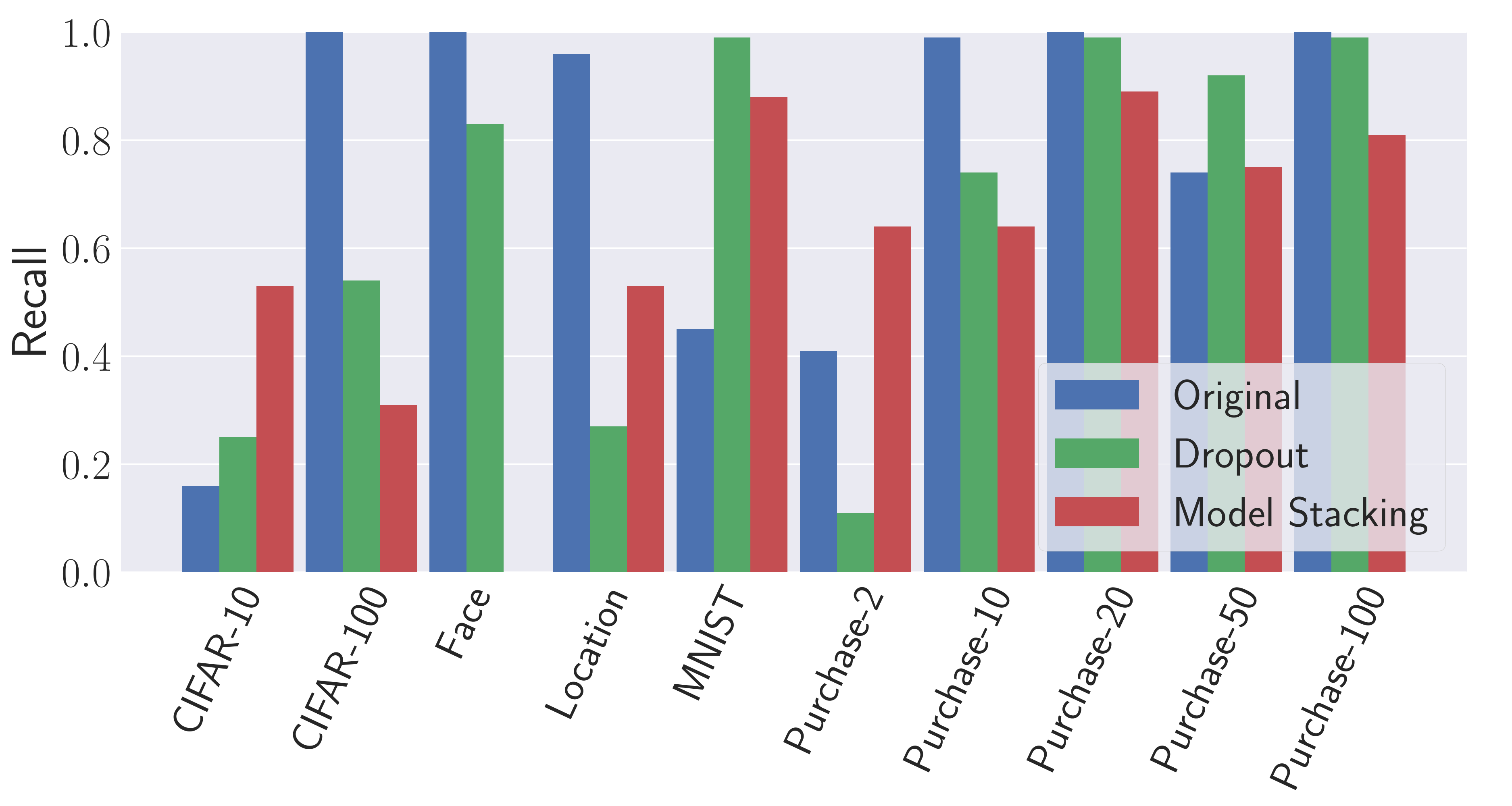}
   \caption{}
   \label{fig:attk3defRecall}
\end{subfigure}
\caption{Comparison of the third adversary's performance 
under both of the defense mechanisms. (a) precision, (b) recall.}
\label{fig:attk3defComp}
\end{figure*}

As the first and second adversaries 
follow the same methodology of building a shadow model (with different assumptions on the dataset), we only show the effectiveness of our defense on the first adversary
as well as on the third adversary to conserve space.
For the first adversary, to fully assess the attack's performance under our defense,
we further assume the attacker knows the defense technique being implemented
and builds her shadow model following the same defense technique.

\begin{figure}
   \includegraphics[width=1\linewidth]{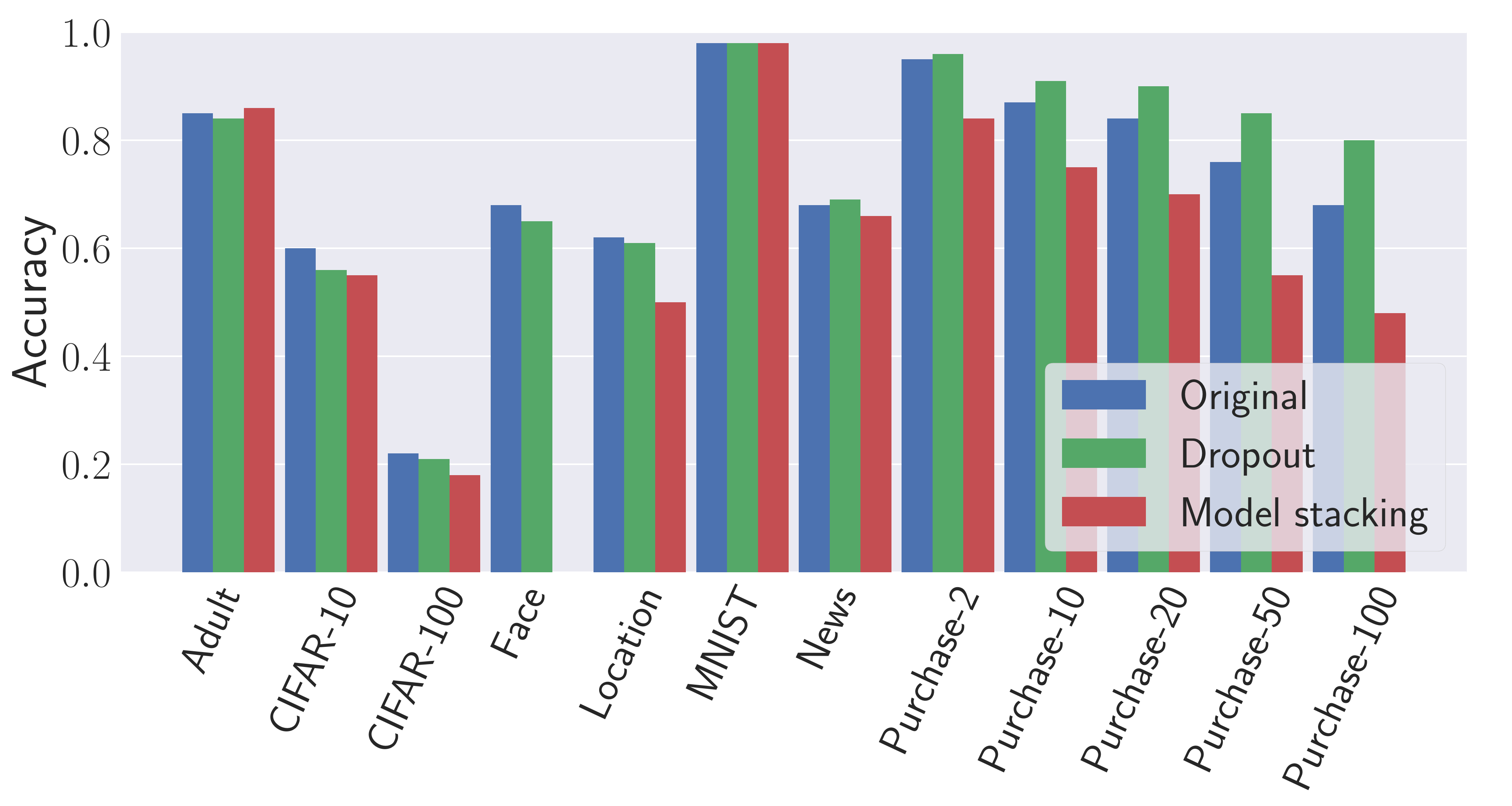}
    \caption{Comparison of the target model's accuracy 
    under both of the defense mechanisms.}
   \label{fig:defAcc} 
\end{figure}

\subsection{Dropout}

\noindent\textbf{Methodology.}
A fully connected neural network contains a large number of parameters
which is prone to overfitting.
Dropout is a very effective method to reduce overfitting
based on empirical evidences.
It is executed by
randomly deleting in each training iteration a fixed proportion (dropout ratio)
of edges in a fully connected neural network model. 
We apply dropout for both the input layer and the hidden layer (see \autoref{sec:attack1})
of the target model.
We set our default dropout ratio to be 0.5.

\begin{figure}[!ht]
\centering
\begin{subfigure}{0.23\textwidth}
   \includegraphics[width=\linewidth]{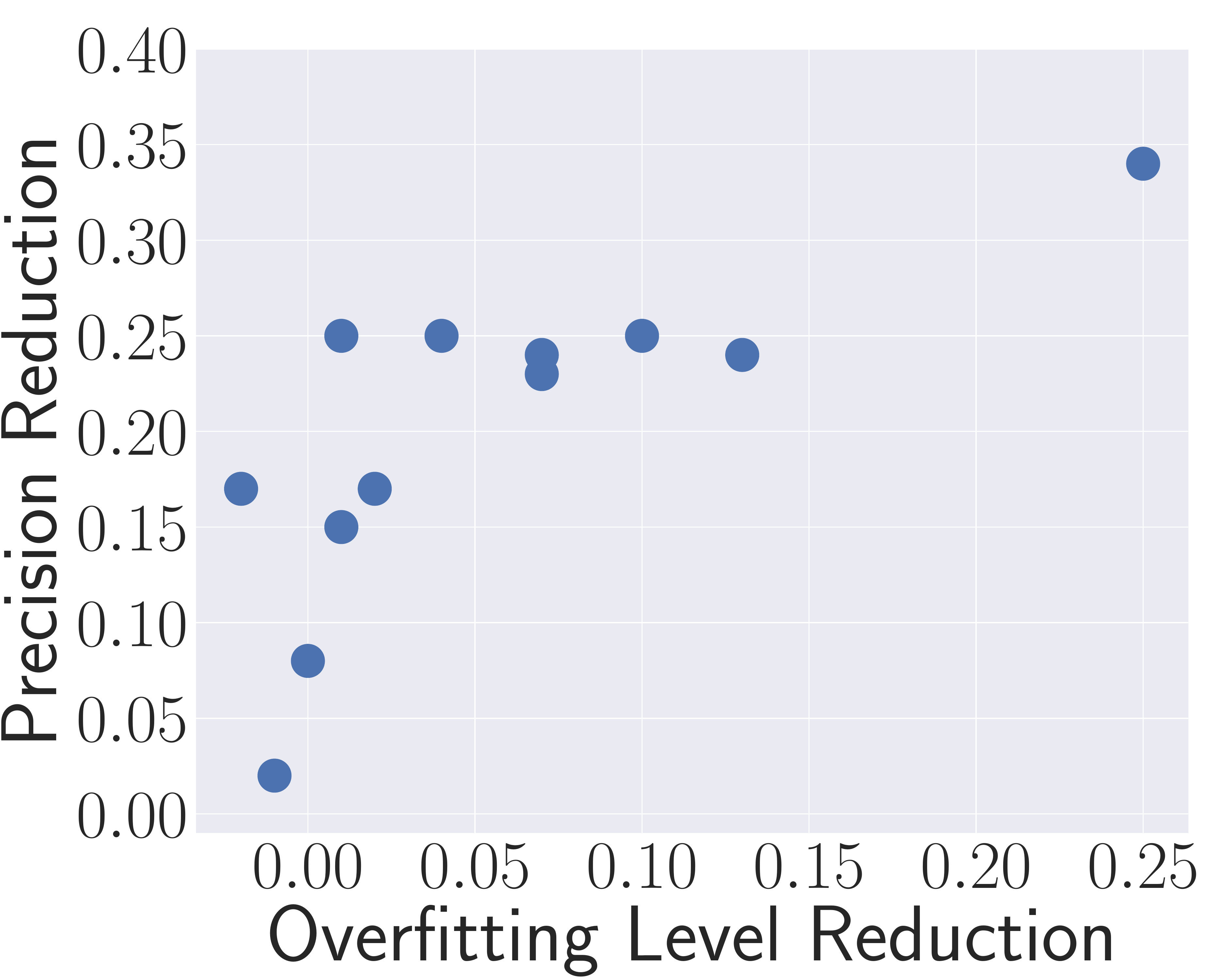}
   \caption{}
   \label{fig:dropoutGapPres} 
\end{subfigure}
\begin{subfigure}{0.23\textwidth}
   \includegraphics[width=\linewidth]{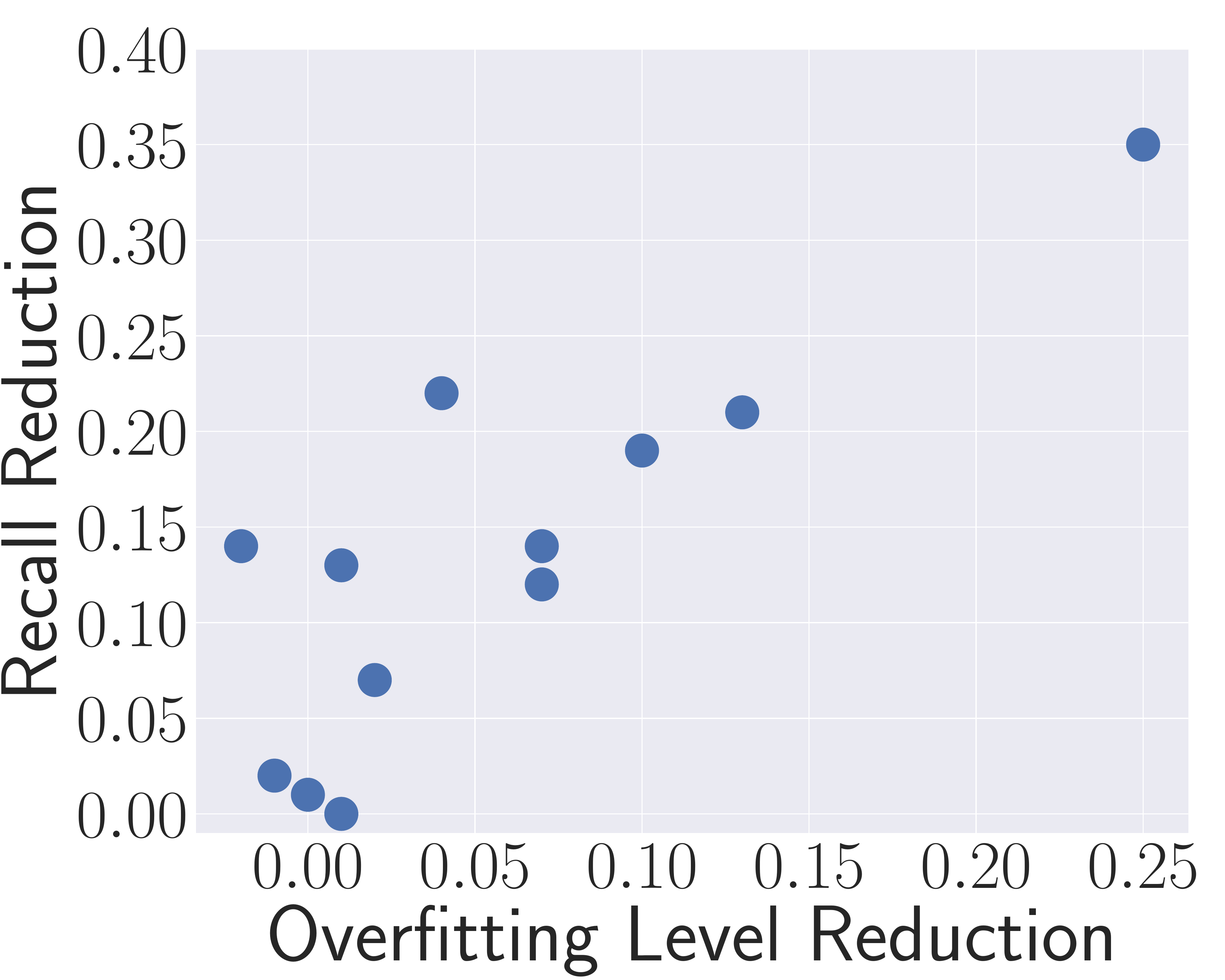}
   \caption{}
   \label{fig:dropoutGapRecall}
\end{subfigure}
\caption{The relation between the overfitting level reduction (x-axis) and the first adversary's performance reduction (y-axis)
when applying dropout as the defense mechanism. (a) precision reduction, (b) recall reduction.}
\label{fig:dropoutGap}
\end{figure}

\begin{figure*}[!ht]
\centering
\begin{subfigure}{0.3\textwidth}
   \includegraphics[width=1\linewidth]{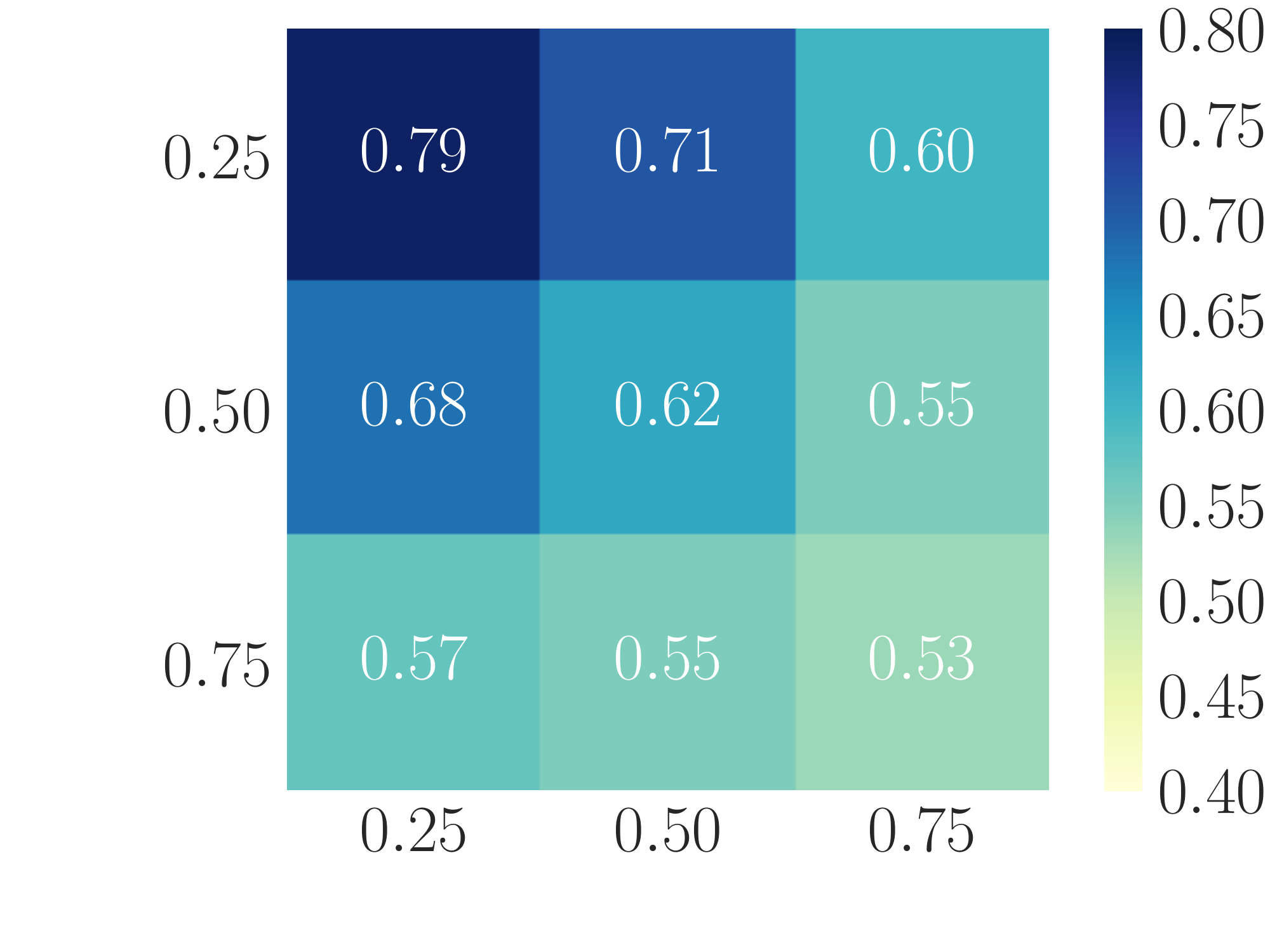}
   \caption{Precision}
   \label{fig:dropoutPrec} 
\end{subfigure}
\begin{subfigure}{0.3\textwidth}
   \includegraphics[width=1\linewidth]{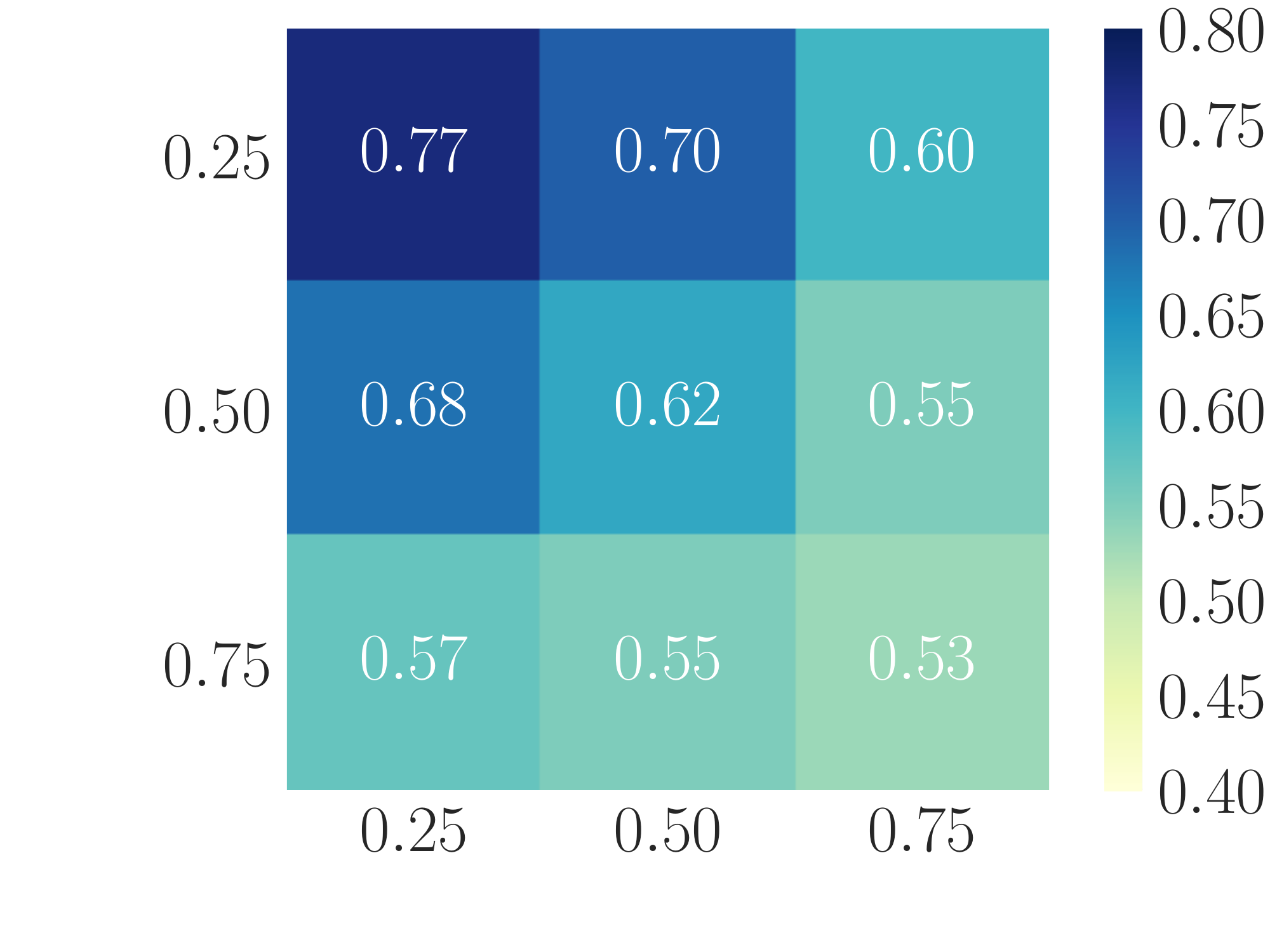}
   \caption{Recall}
   \label{fig:dropoutTransRecall}
\end{subfigure}
\begin{subfigure}{0.3\textwidth}
   \includegraphics[width=1\linewidth]{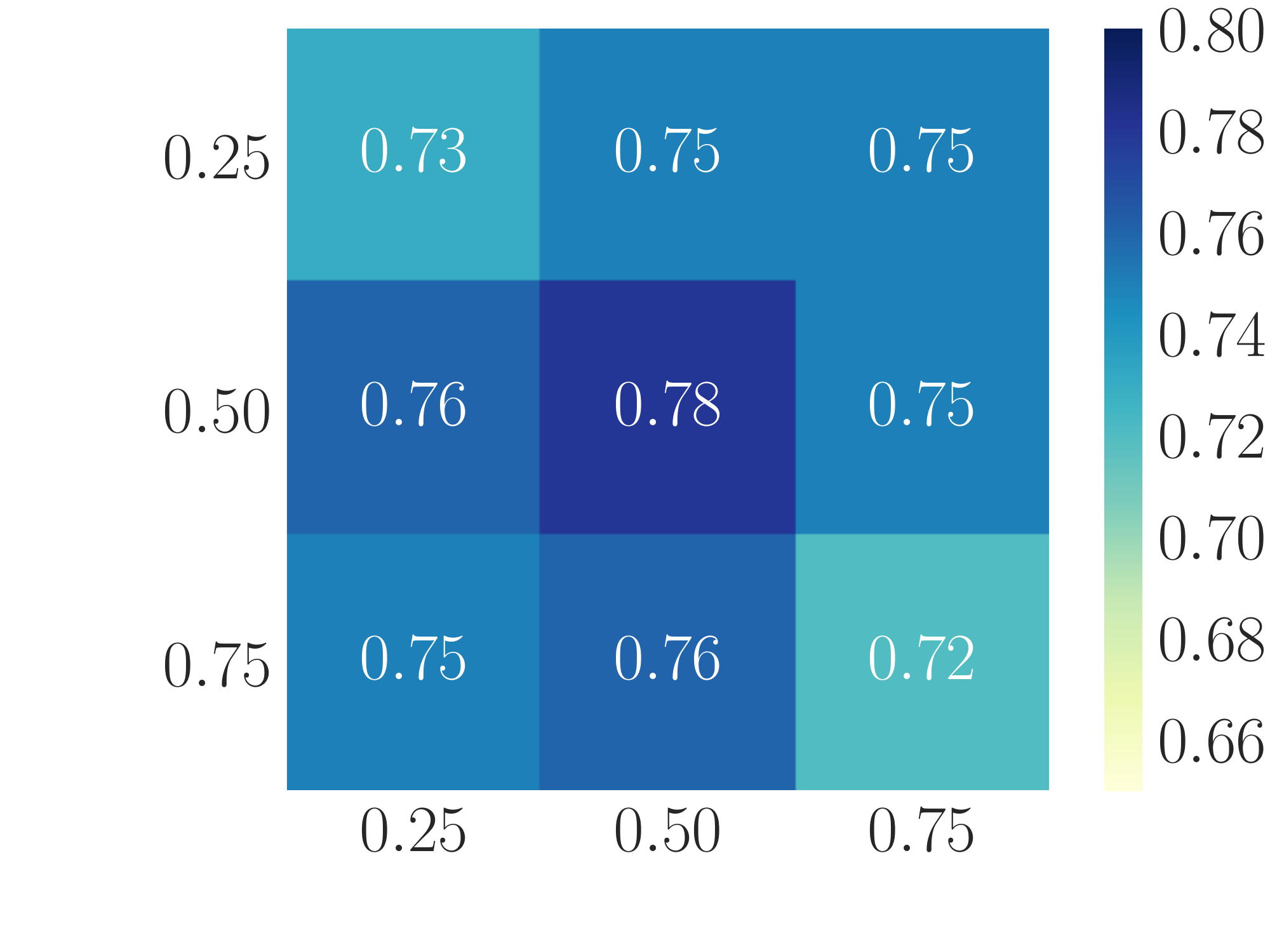}
   \caption{Accuracy}
   \label{fig:dropoutAcc} 
\end{subfigure}
\caption{The effect of the dropout defense on the first adversary's performance, i.e., (a) precision and (b) recall, and on (c) the target model's accuracy
under different dropout ratios in different layers of the neural network.
The x-axis represents the input layer,
and the y-axis represents the hidden layer.}
\label{fig:dropout} 
\end{figure*}

\noindent\textbf{Evaluation.}
We test dropout on all datasets 
against the first adversary and third adversary (except for the News and Adult datasets).
\autoref{fig:defPres} and~\autoref{fig:defRecall} 
compare the performance of the first adversary's performance before and after the dropout defense.
As we can see, the attack performance is reduced in almost all cases.
For instance, the precision of the attack on the Purchase-100 dataset 
drops from 0.89 to 0.64, while the recall drops from 0.86 to 0.63.
In another example, the precision and recall on the CIFAR-100 dataset
drop by more than 30\%.
There is only one case where dropout does not help much,
i.e., the target model trained on the News dataset.

Similarly, the performance of our third adversary is reduced due to dropout (see~\autoref{fig:attk3defComp}).
For example, the precision and recall of the attack on the CIFAR-100 dataset drop by more than 25\% and 40\%, respectively.
However, on some datasets, such as MNIST, 
the recall of the attack even improves.
This indicates that our third adversary is more resistant to dropout than our first adversary.

\autoref{fig:defAcc} further shows the
original target model's performance (prediction accuracy) 
after dropout has been applied.
We observe that, on more than half of the datasets, 
the dropout mechanism even increases the target model's prediction performance.
For instance, on the Purchase-50 dataset, 
the target model's accuracy increases from 0.72 to 0.83.

Figure~\ref{fig:dropoutGap} plots the relation 
between the overfitting level (see~\autoref{sec:attack1}) reduction
and the first adversary's performance reduction after dropout has been applied.
The overfitting level reduction is calculated as the original target model's
overfitting level subtracting the dropout-defended target model's overfitting level.
As we can see, more effective dropout which results in larger reduction on overfitting level
leads to better defense against membership inference attacks.
These results support the argument by Shokri et al.~\cite{SSSS17}
that overfitting is a common enemy for the membership privacy risks and the target model's performance.

So far, we have used 0.5 as the dropout ratio. We further test the effect of varying the dropout ratio of our defense.
We try different dropout ratios on both input and fully connected layers 
while monitoring the results on the first adversary's performance 
and the target model's accuracy.
\autoref{fig:dropout} shows the result on the Purchase-100 dataset.
We first observe that higher dropout ratio leads to lower attack performance.
For instance, dropout ratio 0.75 on both layers reduces the attack's performance to 0.53 precision and recall.
On the other hand,
both large and small dropout ratio result in the low performance of the target model.
This means the accuracy of the target model is the strongest 
when dropout ratio is mediate.
In conclusion, 0.5 dropout ratio is a suitable choice for this defense technique.

\subsection{Model Stacking}

\noindent\textbf{Methodology.}
The dropout technique is effective, 
however, it can only be applied when the target model is a neural network.
To bypass this limitation, we present our second defense technique,
namely model stacking,
which works independently of the used ML classifier.

The intuition behind this defense is that 
if different parts of the target model 
are trained with different subsets of data, 
then the complete model should be less prone to overfitting.
This can be achieved by using ensemble learning.

\begin{figure}[!ht]
\centering
\includegraphics[width=\linewidth]{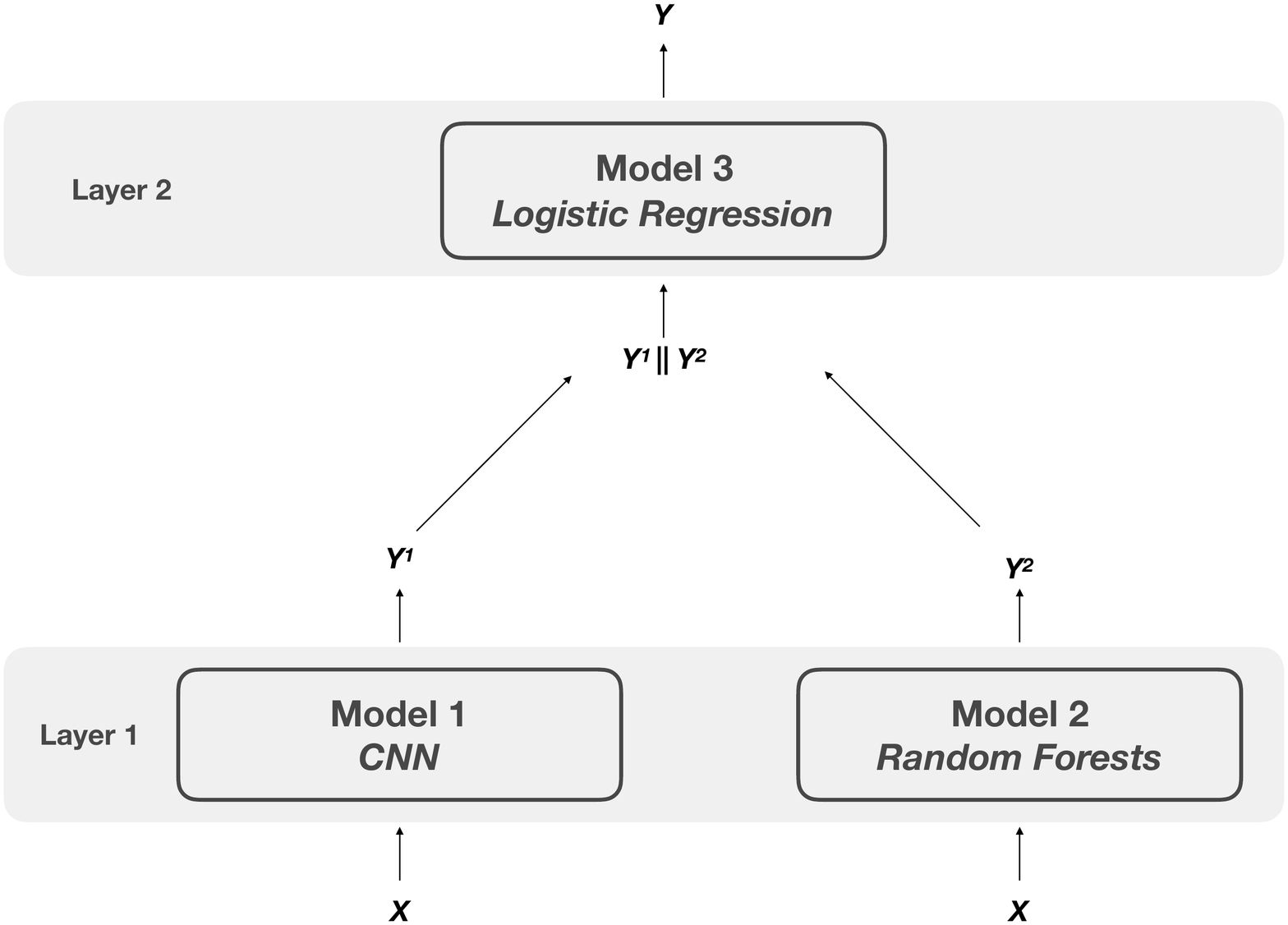}
\caption{The architecture of model stacking.}
\label{fig:defModelStack}
\end{figure}

Ensemble learning is an ML paradigm, where instead of using a single ML model, multiple ML models are combined to construct the final model.
There are different approaches to combine these ML models, such as bagging or boosting.
For our defense, we focus on stacking the models in a hierarchical way.
\autoref{fig:defModelStack} shows a sample architecture for model stacking.

Concretely,
we organize the target model in two layers over three ML models.
The first layer consists of two ML models (the first and second model).
The second layer consists of a single ML model (the third model).
As shown in \autoref{fig:defModelStack},
to get the model's output on some data point $\mathbf{x}$, 
we first apply $\mathbf{x}$ on each of the first two models 
to have their posteriors $\outputvec^1$ and $\outputvec^2$.
We then concatenate both outputs, i.e., $\outputvec^1 || \outputvec^2$,
and apply the result to the third model which predicts the final output $\outputvec$.

To maximize the prevention of overfitting, 
we train the three different models on disjoint sets of data.
The intuition behind is that there is no data point 
seen by more than one model during training.

\noindent\textbf{Evaluation.}
For our evaluation, we use multilayer perceptron or CNN as the first model, random forests as the second model,
and logistic regression as the third model.
We pick this architecture to test the effect of using different machine learning models in the different layers.
However, a different selection of models also suffices.

We build both target and shadow models for the first adversary as described, i.e., each model consists of 3 different ML models.
To train our target and shadow models, we split the data into 12 disjoint sets. 
We use the first 6 sets to train and test our target model, 
and the remaining 6 to train and test the shadow model.

\begin{figure}[!t]
\centering
\begin{subfigure}{0.23\textwidth}
   \includegraphics[width=\linewidth]{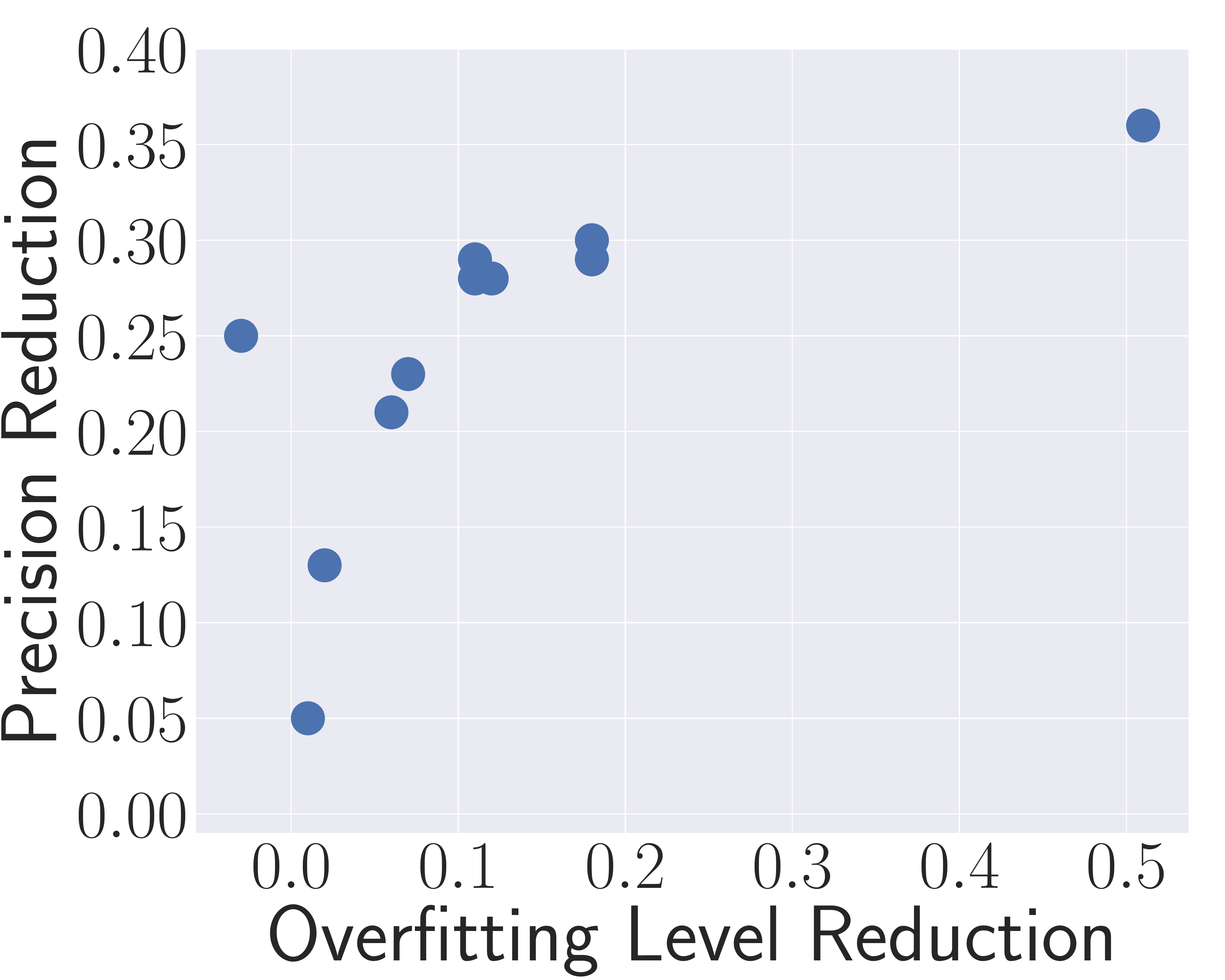}
   \caption{}
   \label{fig:stackingGapPres} 
\end{subfigure}
\begin{subfigure}{0.23\textwidth}
   \includegraphics[width=\linewidth]{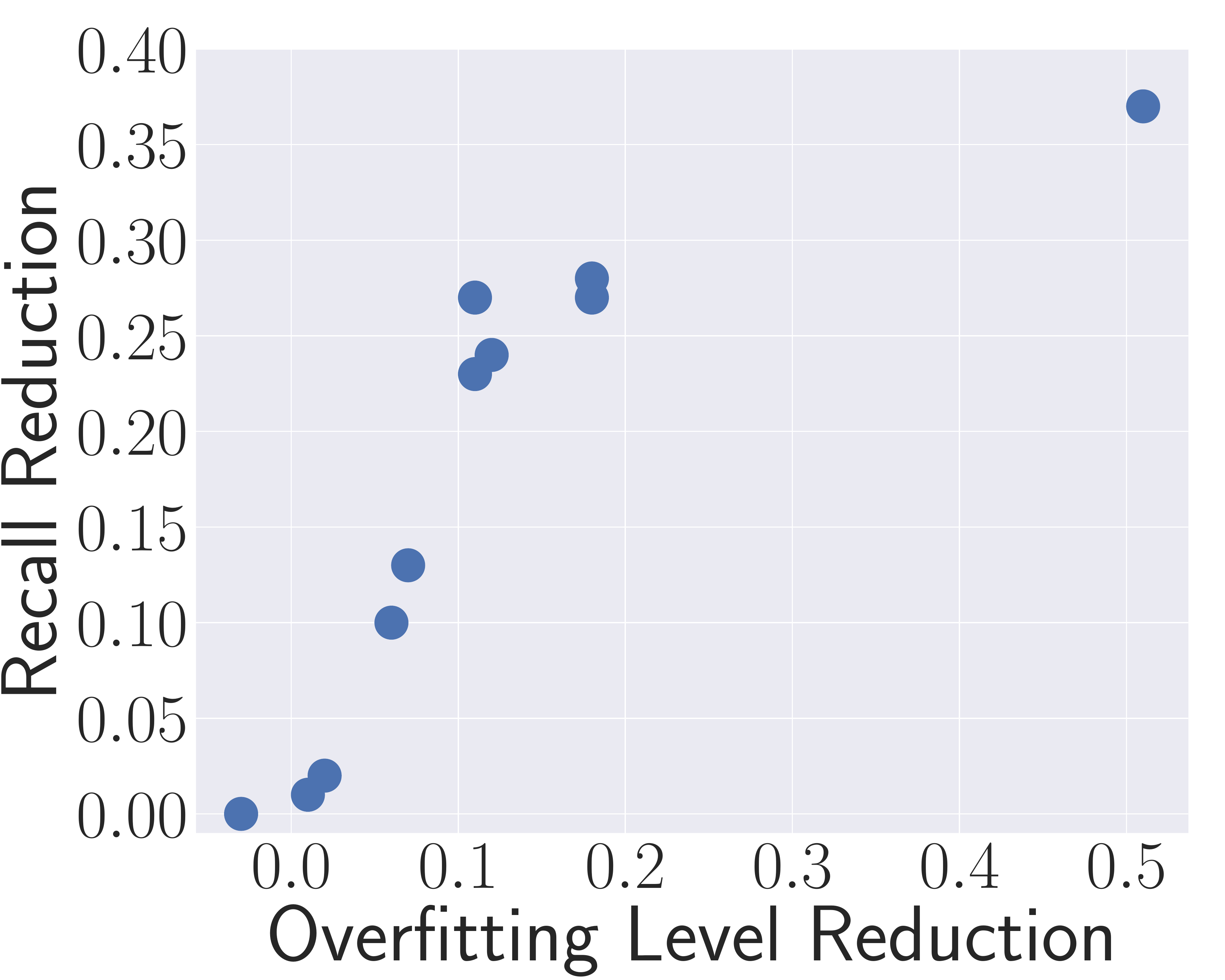}
   \caption{}
   \label{fig:stackingGapRecall}
\end{subfigure}
\caption{The relation between the overfitting level reduction (x-axis) and the first adversary's performance reduction (y-axis)
when applying model stacking as the defense mechanism. (a) precision reduction, (b) recall reduction.}
\label{fig:stackingGap}
\end{figure}

We evaluate this technique on all datasets but the Face dataset as it does not have enough data to provide meaningful results in this setting.
\autoref{fig:defComp} shows the result for the first adversary.
As we can see, model stacking reduces the attack's performance significantly in all cases.
For instance, on the CIFAR-10 dataset,
model stacking reduces the attack's precision and recall by more than 30\%.
Moreover, compared to the dropout defense,
model stacking is more effective in some cases.
Dropout does not change the attack's performance on the News dataset
while model stacking reduces the corresponding precision and recall by 28\%.
The same result can be observed on the Location dataset.
However, model stacking affects target model's accuracy more
than dropout in multiple cases, e.g., the Purchase datasets.
The relation between overfitting level reduction and attack performance reduction
for the model stacking technique is very similar to the one for the dropout technique,
the results are depicted in~\autoref{fig:stackingGap}.

Similarly, the performance of our third adversary drops after model stacking has been applied (see \autoref{fig:attk3defComp}).
For instance, model stacking reduces the attack's performance on the Location dataset by more than 20\% for the precision and 30\% for the recall.
But similar to the dropout defense, exceptions like MNIST also exist.

In conclusion,
if the target model is not based on neural networks,
model stacking is an effective defense technique.
Otherwise, dropout is sufficient to mitigate the membership privacy risks
due to its high utility maintenance.

\section{Related Work}
\label{sec:relwork}

\noindent\textbf{Membership Inference.}
Membership inference attack has been successfully performed in many different data domains,
ranging form biomedical data~\cite{HSRDTMPSNC08,BBHM16}
to mobility traces~\cite{PTC18}.

Homer et al.~\cite{HSRDTMPSNC08} propose the first membership inference attack on genomic data.
This attack relies on the $L_1$ distance between the allele frequencies and the victim's genomic data.
Backes et al.~\cite{BBHM16} generalize this attack to other types of biomedical data.
More recently, Pyrgelis et al.~\cite{PTC18}
have shown that people's aggregate mobility traces are also prone to membership inference attack.
They first formalize membership inference as a distinguishability game.
Then, they implement the attack with machine learning classifiers.
Large-scale evaluation on two real-world datasets 
has demonstrated their attack's effectiveness.
Moreover, the authors
show their framework 
can easily incorporate different defense mechanisms, such as differential privacy,
to allow a comprehensive evaluation of membership inference risks.

\noindent\textbf{Membership Inference Against Machine Learning.}
Shokri et al.~\cite{SSSS17} present the first membership inference attack against machine learning models.
The key contribution of this work is the proposal of shadow model training,
which aims at mimicking the target model's behavior to generate training data for the attack model.

The first adversary in the current paper follows a very similar setting.
We have shown that one shadow model and one attack model
are sufficient to achieve an effective attack compared to the proposal of multiple shadow models and attack models by Shokri et al.~\cite{SSSS17}.
Moreover, we show that data transferring attack can bypass 
the expensive synthetic data generation scheme
and achieve a very similar performance.
Another major contribution of our paper is the two effective defense mechanisms,
such as dropout and model stacking.
Many recent works 
have studied membership inference 
against machine learning as well
from different angles~\cite{LBG17,HMDC17,YGFJ18,LBWBWTGC18}.

\noindent\textbf{Attacks Against Machine Learning.}
Besides membership inference,
there exist multiple other types of attacks against ML models.
Fredrikson et al.~\cite{FLJLPR14} present the model inversion attack 
in biomedical data setting.
In this scenario,
an attacker aims to infer the missing attributes of her victim,
relying on the output of a trained ML model.
Later, model inversion attack is generalized
to a broader scenario~\cite{FJR15}.
For instance, the authors show that it is feasible for an attacker to reconstruct 
a recognizable face of her victim with model inversion.

Tram\`er et al.~\cite{TZJRR16} propose another attack on ML models, namely model extraction attack.
This attack aims at stealing the ML model, i.e., the model's learned parameters,
through the output of MLaaS API itself.
They first propose an equation-solving attack,
where an attacker queries MLaaS API multiple times and use the output posteriors
to construct a set of equations.
By solving these equations, the attacker can obtain the weight of the ML model.
Tram\`er et al.~\cite{TZJRR16} further propose a path-finding algorithm,
which is the first practical method to steal decision trees.
In the end, Tram\`er et al. show that even ML models
which do not provide prediction posteriors
but only prediction class labels
can still be stolen with 
retraining strategies, such as active learning.
It is worth noting that
due to the effectiveness of the model extraction attack,
we do not consider hiding posteriors as one valid defense mechanism 
in this paper.

Another major family of attacks against machine learning
is adversarial examples~\cite{PMGJCS17,VL14,CW17,LV15,TKPGBM17,PMJFCS16,XEQ18}.
In this setting,
an attacker adds a controlled amount of noise to a data point
which aims to fool a trained ML model to mis-classify the data point.
Adversarial examples can cause severe risks in multiple domains,
such as autonomous driving, and voice recognition.
On the other hand, researchers have recently 
shown that adversarial examples
can also help to protect users' privacy in online social networks~\cite{OFS17,JG18,ZHRLPB18}.

\noindent\textbf{Privacy-Preserving Machine Learning.}
Another relevant line of work is privacy-preserving machine learning~\cite{LJLA17,HMEK11,EFGKLT09,DGLLNW16,GSBRDZE17,MZ17,BIKMMPRSS17,BPTG15,BBBEHHL17}.
Mohassel and Zhang~\cite{MZ17}
present efficient protocols for training
linear regression, logistic regression, and neural networks
in a privacy-preserving manner.
Their protocols fall in the two-server mode
where data is distributed over two non-colluding servers.
The authors use two-party computation to implement these protocols.
Bonawitz et al.~\cite{BIKMMPRSS17}
propose a protocol for secure aggregation over high-dimensional data,
which is a key component for distributed machine learning.
The protocol is also based on multi-party computation,
and the authors show its security under both honest-but-curious
and active adversary setting.
Large-scale evaluation demonstrates the efficiency
of this protocol.

Besides privacy-preserving model training,
other works study privacy-preserving classification.
Bost et al.~\cite{BPTG15}
design three protocols based on homomorphic encryption.
They concentrate on three ML classifiers 
including hyperplane decision, Naive Bayes, and decision trees,
and show that their protocols can be efficiently executed.
Based on the scheme of Bost et al.,
Backes et al.~\cite{BBBEHHL17}
build a privacy-preserving random forests classifier
for medical diagnosis.
Besides the above,
many recent works have tackled security and privacy in machine learning from various perspectives~\cite{SRS17,HZGSABF18,CXXLBKZ18,RRK17,CG17,MSCS18,HSSSW18,WG18,RCK18,JOBLNL18,SMKID18,NSH18,TB18,HSSSW18}.
\section{Conclusion}
\label{sec:conclu}

Training data is a key factor that drives machine learning model being widely adopted in real-world applications.
However, ML models suffer from membership privacy risks.
The existing membership inference attacks have shown effective performance,
but their applicability is limited due to strong assumptions
on the threat model.
In this paper, we gradually relax these assumptions
towards a more broadly applicable attack scenario.

Our {\bf first adversary} utilizes only one shadow model.
Extensive experiments show that this attack
achieves a very similar performance as the previous one
which utilizes multiple shadow models.
As shadow models are established through MLaaS,
our proposal notably reduces the cost of conducting the attack.
We further perform the combining attack
which does not require knowledge of the type of classifiers
used in the target model.

The attack assumption is further relaxed for the {\bf second adversary},
who does not have access to a dataset that comes from 
the same distribution as the training data of the target model.
This is a more realistic attack scenario,
but the previously proposed synthetic data generation solution
can only be applied in specific cases.
In contrast, we propose data transferring attacks,
where the adversary utilizes another dataset to build a shadow model
and generates the corresponding data to attack the target model.
Through experiments,
we have discovered that data transferring attack also achieves strong membership inference while being more general, realistic and widely applicable.

The {\bf third adversary} has a minimal set of assumptions,
i.e., she does not need to construct any shadow model
and her attack is performed in an unsupervised way.
We show that even in such a simple setting,
membership inference is still effective.

Our {\bf evaluation} is comprehensive and fully demonstrates
the severe threat of membership privacy in ML models under those generalized conditions on 8 diverse datasets.

To remedy the situation,
we propose {\bf two defense mechanisms}.
As we show the connection between overfitting and sensitivity to membership inference attacks, we investigate techniques that are designed to reduce overfitting.
The first one, namely dropout, randomly deletes a certain proportion of edges 
in each training iteration
in a fully connected neural network,
while the second approach, namely model stacking,
organizes multiple ML models in a hierarchical way.
Extensive evaluation shows that indeed our defense techniques
are able to largely reduce membership inference attack's performance,
while maintaining a high-level utility, i.e., high target model's prediction accuracy.
\section*{Acknowledgments}
The research leading to these results has received funding from the European Research Council under the European Union's Seventh Framework Programme (FP7/2007-2013)/ ERC grant agreement no. 610150-imPACT.

\balance
\bibliographystyle{IEEEtranS}
\bibliography{normal_generated}

\end{document}